\documentclass[12pt]{article}
\usepackage{amsfonts,amssymb,epsfig}
\usepackage[latin1]{inputenc}
\usepackage[english]{babel}
\usepackage{hyperref}
\usepackage{color}
\newtheorem{thm}{Th\'eor\`eme}[section]
\newtheorem{cor}[thm]{Corollaire}
\newtheorem{lem}[thm]{Lemme}
\newtheorem{pro}[thm]{Proposition}
\newtheorem{dfn}[thm]{D\'efinition}
\newtheorem{rmk}[thm]{Remark}
\newtheorem{expl}[thm]{Exemple}

\def\dessous#1\sous#2{\mathrel{\mathop{\kern0pt#2}\limits_{#1}}}

\newcommand{\R}{\mathbb R}
\newcommand{\C}{\mathbb C}

\newcommand{\beq}{\begin{eqnarray}}
\newcommand{\eeq}{\end{eqnarray}}
\newcommand{\bpro}{\begin{pro}}
\newcommand{\epro}{\end{pro}}
\newcommand{\blem}{\begin{lem}}
\newcommand{\elem}{\end{lem}}
\newcommand{\bdfn}{\begin{dfn}}
\newcommand{\edfn}{\end{dfn}}
\newcommand{\bcor}{\begin{cor}}
\newcommand{\ecor}{\end{cor}}
\newcommand{\bthm}{\begin{thm}}
\newcommand{\ethm}{\end{thm}}
\newcommand{\bex}{\begin{expl}}
\newcommand{\eex}{\end{expl}}
\newcommand{\brmk}{\begin{rmk}}
\newcommand{\ermk}{\end{rmk}}
\newcommand{\benum}{\begin{enumerate}}
\newcommand{\eenum}{\end{enumerate}}
\newcommand{\bitem}{\begin{itemize}}
\newcommand{\eitem}{\end{itemize}}

\begin{document}
\begin{center}
{\Large\bf {Unitary maps on Hamiltonians of an electron moving in a plane and coherent states construction}}\\
\vspace{0.5cm}
Isiaka Aremua$^{1,3}$ and Laure Gouba$^{2,3}$ \\
$^{1}${\em
Universit\'{e} de Lom\'{e} (UL), Facult\'{e} Des Sciences  (FDS), D\'{e}partement de Physique} \\
{\em Laboratoire de Physique des Mat\'eriaux et des Composants \`a Semi-Conducteurs}\\
{\em Universit\'{e} de Lom\'{e} (UL), 01 B.P. 1515  Lom\'{e} 01, Togo.\\
 E-mail: claudisak@gmail.com}\\  
$^{2}${\em
The Abdus Salam International Centre for Theoretical Physics (ICTP),\\
Strada Costiera 11, I-34151 Trieste Italy.\\
 E-mail: laure.gouba@gmail.com}\\
 $^{3}$International Chair of Mathematical Physics
 and Applications. \\
 {\em ICMPA-UNESCO Chair,  University of Abomey-Calavi}\\
 {\em 072 B.P. 50 Cotonou, Republic of Benin.}\\
\vspace{1.0cm}

\today

\begin{abstract}
\noindent
In this work we consider a model of an electron moving in a 
plane under uniform external magnetic and electric fields.
We inverstigate the action of unitary maps on the associated 
quantum Hamiltonians and construct the coherent states of 
Gazeau-Klauder type. 
\end{abstract}

\end{center}
\setcounter{footnote}{0}

\section{Introduction}

Coherent states (CSs) were introduced for the first time by Schr\"{o}dinger in 1926 \cite{schrodinger} in his study
of quantum states that restore the classical behavior of a quantum observable. For various generalizations, approaches and their properties one may consult 
\cite{klauder-skagerstam}\cite{perelomov}\cite{gazbook09}\cite{ali-antoine-gazeau}\cite{monique} and references therein.
 CSs, known as an overcomplete family of vectors, represent one
of the most fundamental framework for the analysis, or decomposition, of states in the Hilbert spaces, which are the underlying mathematical structures of several physical phenomena. 

The system of charged quantum particles interacting with a constant
magnetic field continues to attract intensive studies and represents one of
the most investigated quantum systems, mainly motivated by condensed
matter physics and quantum optics. Some recent works discuss this quantum system and its related different kind of associated coherent states \cite{ab1} \cite{abh} \cite{ aremuaromp} \cite{dodonov1} \cite{aremua-gouba1} \cite{aremua-gouba2}.  

It has been established in \cite{ab1}\cite{abh} \cite{aremuahbs}\cite{aremuajmp} that when dealing with the Hilbert-Schmidt operators in the case of Tomita-Takesaki modular structures \cite{takesaki1}, these latters are associated to both directions of the external magnetic field. Then, it is possible via the Wigner transform to obtain the equivalent of  Hilbert-Schmidt operators on $\mathcal B_2(\mathfrak H)$, $\mathfrak H$ being the quantum Hilbert space of the harmonic oscillator in the Schr\"{o}dinger representation, as operators  acting on $L^2(\R^2, dxdy)$ the Hilbert space associated to the quantum Hamiltonians describing  the electron system. 

In \cite{aremuajmp},  the Glauber-Sudarshan-$P$-representation of the density operator has been used to study harmonic oscillator quantum systems and models of spinless electrons moving in a two-dimensional noncommutative space, subject to a magnetic field background coupled with a harmonic oscillator. Relevant statistical properties such as the $Q$-Husimi distribution and the Wehrl entropy have been investigated. Recently, in \cite{aremuaromp1}, the  density operator representation in the context of multi-matrix vector coherent states basis
is performed and applied to Landau levels of an electron in an electromagnetic field coupled to an isotropic harmonic potential. Main relevant statistical properties such as the Mandel $Q$-parameter and the signal-to-quantum-noise ratio have been  derived and discussed.

In  our recent works \cite{aremua-gouba1}\cite{aremua-gouba2}, we constructed CSs for a system of an electron moving in a plane under uniform external magnetic and electric fields, first in the context of discrete and continuous spectra, and next by considering both spectra purely discrete. These coherent states obey the Gazeau-Klauder criteria \cite{gazeau-klauder} that a family of CSs must satisfy. Inspired by the Tomita-Takesaki modular theory investigated on the  Hilbert-Schmidt  operators Hilbert space through  the Wigner map \cite{takesaki1} and similar discussions in \cite{abh}, we pursue the discussions of the works in \cite{aremua-gouba1, aremua-gouba2} as follows.
We consider some unitary transforms defined on the Hilbert space $\mathcal B_{2}(\mathfrak H)$ of Hilbert-Schmidt operators, with $\mathfrak{H} = L^2(\R)$, the Hilbert space in the Schr\"{o}dinger representation. Then, we explore the operator representations through these maps. The CSs of the Gazeau-Klauder type  are therefore built and satisfy all Klauder's minimal requirements for CSs. Besides, using unitary operators realizing unitary irreducible representations of the Weyl-Heisenberg group, GK-CSs are also achieved.

The paper is organized as follows. Section \ref{sec2} gives some preliminaries on  quantum Hamiltonians describing the electron moving in a plane under uniform external magnetic and electric fields, also on the Wigner transform and its inverse. In Section \ref{sec3}, we discuss the operator representations related to the unitary transforms. Gazeau-Klauder CSs (GK-CSs) are constructed in Section \ref{sec4}, with Klauder's minimal requirements investigated. Besides, using unitary operators  realizing unitary irreducible representations of the Weyl-Heisenberg group, GK-CSs are also provided. Concluding remarks are given in section \ref{sec5}.
 
\section{Preliminaries}\label{sec2}
In this section, we give a summary of the physical model  of an electron
moving in a plane under uniform external magnetic and electric fields related quantum Hamiltonians presented in \cite{aremua-gouba1} in both symmetric gauges, then we provide some basic notions on the Wigner transform and its inverse as it has been introduced in \cite{ali-antoine-gazeau} \cite{ab1} \cite{abh} \cite{aremuahbs}.

\subsection{Quantum Hamiltonians in both symmetric gauges}

We consider an electron moving in a  plane $(x,y)$ in the uniform external
electric field  $\overrightarrow{E} =-\overrightarrow{\nabla}\Phi(x,y)$ and the uniform external magnetic field $\overrightarrow{B}$ which is perpendicular to the plane  described by the Hamiltonian
\begin{equation}{\label{es0}}
H = \frac{1}{2M}\left(\overrightarrow{p} +
\frac{e}{c}\overrightarrow{A}\right)^{2} - e\Phi\;,
\end{equation}
where $\overrightarrow{A}$ is the the magnetic vector potential.
According to our study of the same model in reference \cite{aremua-gouba1}, we have the following summary.

In the symmetric gauge $\overrightarrow{A} = \left(\frac{B}{2}y, -\frac{B}{2}x \right)$ with  the scalar potentiel given by $ \Phi(x,y) = -Ey$, 
the corresponding  classical Hamiltonian, obtained from (\ref{es0}), denoted by $
H_{1}$, reads 
\begin{equation}{\label{es1}} 
H_{1}(x,y,p_x,p_y)   =
\frac{1}{2M}\left[\left(p_{x} + \frac{eB}{2c}y\right)^{2} +
\left(p_{y} - \frac{eB}{2c}x\right)^{2} \right] + eEy, 
\end{equation}
where $x, y, p_x, p_y$ are the canonical classical variables.  Performing canonical quantization, the classical variables are promoted respectively to the operators $\hat X, \hat Y, \hat P_x, \hat P_y$ and the corresponding Hamiltonian operator is given by
\begin{equation}{\label{esh}} 
H_{1}(X,Y,P_x,P_y)   =
\frac{1}{2M}\left[\left(P_{x} + \frac{eB}{2c}Y\right)^{2} +
\left(P_{y} - \frac{eB}{2c}X\right)^{2} \right] + eEY.
\end{equation}
Through some change of variables, where the details are given in \cite{aremua-gouba1} the Hamiltonian operator in (\ref{esh}) is splitted into two commuting  parts as follows:
\begin{equation}\label{es7}
\hat H_1 = \frac{1}{4M}\left(b^{\dag}b + bb^{\dag}\right) - \frac{\lambda}{2M}\left(d^{\dag} + d\right) - \frac{\lambda^{2}}{2M}.
\end{equation}

\begin{equation}\label{quah000}
\hat H_{1} = \hat H_{1_{OSC}} -  \hat T_1, 
\end{equation}
where  $\hat H_{1_{OSC}}$ denotes the harmonic oscillator part 
\begin{equation}\label{equah1osc}
\hat H_{1_{OSC}} = \frac{1}{4M}(b^{\dag}b + bb^{\dag}),
\end{equation} 
while the
part linear in $d$ and  $d^{\dag}$ is given by 
\begin{equation}
\label{tfunc1}
\hat T_1 =
\frac{\lambda}{2M}(d^{\dag} + d) + \frac{\lambda^{2}}{2M}. 
\end{equation} 
The following commutation relations 
\begin{equation}\label{commuta000}
[b, b^{\dag}] = 2M\hbar \omega_c \mathbb I_{\mathfrak H_{s}}, \qquad [d^{\dag}, d] = 2M\hbar \omega_c \mathbb I_{\mathfrak H_{s}}
\end{equation}
 where $\mathbb  I_{\mathfrak H_{s}}$ is the identity operator in $\mathfrak H_{s} = L^{2}(\R^{2}, dxdy)$ are satisfied. Therefore,
the eigenvectors and the energy spectrum of
the Hamiltonian $\hat H_{1}$ are determined by the following formulas:
\begin{eqnarray}{\label{es17}} 
\Psi_{n, \alpha} &=& \Phi_{n} \otimes
\phi_{\alpha} \equiv |n,\alpha\rangle, \quad \phi_{\alpha} \equiv\phi_{\alpha}(x,y) =
e^{i (\alpha x + \frac{M \omega_{c}}{2 \hbar}xy)}, \;\;\;
\alpha \in \R, \cr \cr \mathcal
E_{(n,\alpha)} &=& \frac{\hbar \omega_{c}}{2}(2n + 1) - \frac{\hbar
\lambda}{M}\alpha - \frac{\lambda^{2}}{2M},  \;\;\; \;\; n= 0, 1, 2,
\dots. 
\end{eqnarray}
In the symmetric gauge $\overrightarrow{A} = \left(-\frac{B}{2}y,\frac{B}{2}x \right)$ with the scalar potential  given by\- $\Phi(x,y) = -E x$,
the classical Hamiltonian $H$ in equation  (\ref{es0}) becomes
\begin{equation}{\label{ei19}}
H_{2}(x,y,p_x,p_y)  =
\frac{1}{2M}\left[\left(p_{x} - \frac{eB}{2c}y\right)^{2} +
\left(p_{y} + \frac{eB}{2c}x\right)^{2}\right] + eEx,
\end{equation} 
and its corresponding Hamiltonian operator is given by
\begin{equation}{\label{ei19op}}
\hat H_{2}(X,Y,P_x,P_y)  =
\frac{1}{2M}\left[\left(P_{x} - \frac{eB}{2c}Y\right)^{2} +
\left(P_{y} + \frac{eB}{2c}X\right)^{2}\right] + eEX,
\end{equation}
According to the changes of variables where details are given in \cite{aremua-gouba1}, the Hamiltonian operator $\hat H_{2}$ in equation (\ref{ei19op})  
can be then written as
\begin{equation}{\label{ei24}} 
\hat H_{2} =
\frac{1}{4M}(\mathfrak b^{\dag}\mathfrak b + \mathfrak b\mathfrak b^{\dag}) - \frac{\lambda}{2M}(\mathfrak d^{\dag} +
\mathfrak d) - \frac{\lambda^{2}}{2M},
\end{equation} 
with 
\begin{equation}\label{eqsop001}
[\mathfrak b, \mathfrak b^{\dag}] = 2M\hbar \omega_c \mathbb I_{\mathfrak H_{s}}, \qquad [\mathfrak  d^{\dag}, \mathfrak  d] = 2M\hbar \omega_c \mathbb I_{\mathfrak H_{s}},
\end{equation}
where the harmonic oscillator part is given by 
\begin{equation}\label{equah2osc}
\hat H_{2_{OSC}} = \frac{1}{4M}(\mathfrak b^{\dag}\mathfrak b + \mathfrak b\mathfrak b^{\dag})
\end{equation}
and the linear part by 
\begin{equation}\label{linp001}
\hat T_{2} = \frac{\lambda}{2M}(\mathfrak d^{\dag} +
\mathfrak d) + \frac{\lambda^{2}}{2M}.
\end{equation}
In addition, we also have
\begin{equation}
[b, \mathfrak b^{\dag}] = 0 = [\mathfrak b, b^{\dag}] \quad \mbox{and} \quad 
[d, \mathfrak d^{\dag}] = 0 = [\mathfrak d, d^{\dag}] .
\end{equation}
The eigenvectors and the eigenvalues of the
Hamiltonian $\hat H_{2}$, as previously determined for $\hat H_{1}$, are obtained  as 
\begin{eqnarray}{\label{eig003}} 
\Psi_{l, \alpha} &=& \Phi_{l} \otimes
\phi_{\alpha} \equiv |l,\alpha\rangle, \quad \phi_{\alpha} \equiv \phi_{\alpha}(x,y) = e^{i (\alpha y +
\frac{M \omega_{c}}{2 \hbar}xy)} \;\;\; \alpha \in \R,\cr \cr \mathcal
E_{(l,\alpha)} &=& \frac{\hbar \omega_{c}}{2}(2l + 1) - \frac{\hbar
\lambda}{M}\alpha - \frac{\lambda^{2}}{2M} \;\;\; \;\; l= 0, 1, 2,
\dots. 
\end{eqnarray}
The eigenvectors denoted $|\Psi_{nl}\rangle : = |\Phi_n\rangle \otimes |\Phi_l\rangle$ of $\hat H_{1_{OSC}} $ and spanning the Hilbert space $\mathfrak H_{s} = L^{2}(\R^{2}, dxdy)$,  can be so chosen that they are also
the eigenvectors of $\hat H_{2_{OSC}}$, since $[\hat H_{1_{OSC}}, \hat H_{2_{OSC}}] = 0, $  as follows:
\begin{equation}{\label{equa37}}
\hat H_{1_{OSC}}|\Psi_{nl}\rangle  = \hbar \omega_{c} \left(n + \frac{1}{2}\right)
|\Psi_{nl}\rangle, \,\,  \hat H_{2_{OSC}}|\Psi_{nl}\rangle  = \hbar
\omega_{c} \left(l + \frac{1}{2}\right) |\Psi_{nl}\rangle, \; n, l  = 0,1,2,\dots
\end{equation}
so that $\hat H_{2_{OSC}}$ lifts the degeneracy of $\hat H_{1_{OSC}}$ and vice versa.

\subsection{The Wigner transform }
Let $\mathcal B_{2}(\mathfrak H)\simeq \mathfrak H \otimes \bar{\mathfrak H}$  be the Hilbert space of Hilbert-Schmidt operators on $\mathfrak H = L^{2}(\R)$, spanned by the eigenvectors (\ref{es17}) and (\ref{eig003}) of $\hat H_{OSC} = \frac{\omega_c}{2}(\mathcal Q^2 + \mathcal P^2), \; [\mathcal Q,  \mathcal P] = \imath \mathbb I_{\mathfrak H}$, corresponding to both Hamiltonians $\hat H_{1_{OSC}} = \frac{\omega_c}{2}(\mathcal Q^2_1 + \mathcal P^2_1)$ or $\hat H_{2_{OSC}}= \frac{\omega_c}{2}(\mathcal Q^2_2 + \mathcal P^2_2)$, where for e.g,  
$\mathcal Q_1 = \frac{1}{2\sqrt{M\omega_c \hbar}}(b + b^{\dag}), \, \mathcal P_1 = \frac{i}{2\sqrt{M\omega_c\hbar}}(b^{\dag} - b)$ (see Eq. (\ref{commuta000}))  with $[\mathcal  Q_1,  \mathcal P_1] = \imath \mathbb I_{\mathfrak H_{s}}$,
in the Schr\"{o}dinger representation. The $\mathcal B_{2}(\mathfrak H)$  basis
vectors are given by
\begin{eqnarray}{\label{equa43}}
\Phi_{nl} := |\Phi_{n}\rangle \langle \Phi_{l}|, \quad n, l= 0,1,2,\dots, \infty.
\end{eqnarray}
Consider the unitary map
$U(x,y)$ on $\mathcal B_{2}(\mathfrak H)$ given by
\begin{eqnarray}\label{unitp001}
(U(x,y)\Phi)(\xi) =  e^{-\imath x \left(\xi - y/2 \right)}\Phi\left(\xi -
y \right),
\end{eqnarray}
with  $U(x,y) = e^{-\imath (xQ + y P)}$,   $Q$ and $P$ being the usual position and momentum operators in the Schr\"{o}dinger 
representation satisfying $[Q, P] = \imath \mathbb I_{\mathfrak H_{s}}$. 
 
Next, if $A$ and $B$ are two operators on $\mathfrak H$, the operator $A \vee B$ is defined  by
\begin{eqnarray}
A \vee B(X) = AXB^{*}, \, X \in \mathcal B_{2}(\mathfrak H).
\end{eqnarray}
Given any vector $X \in \mathcal B_{2}(\mathfrak H)$,
$X = |\Phi \rangle \langle \Psi|$, one has
\begin{eqnarray}{\label{map1}}
 && \mathcal W: \mathcal B_{2}(\mathfrak H) \rightarrow L^{2}(\R^{2}, dxdy) \cr
 \cr
(\mathcal WX)(x,y) &:=& \frac{1}{(2\pi)^{1/2}} Tr\left[U(x,y)^{*}X\right] = \frac{1}{(2\pi)^{1/2}} \langle U(x,y)\Psi|\Phi \rangle_{\mathfrak H} \cr
&=& \frac{1}{(2\pi)^{1/2}}\int_{\R}e^{\imath x\left(\xi -
y/2 \right)}\overline{\Psi\left(\xi- y \right)}\Phi(\xi)d\xi.
\end{eqnarray}
The mapping $\mathcal W$ is often referred to as the Wigner transform in the physical literature and it is well known to be unitary \cite{ali-antoine-gazeau}

\subsection{Inverse of the Wigner transform}
Let us determine the inverse of the map $\mathcal W$ on   the Hilbert space $L^{2}(\R^{2}, dxdy)$ where the group $G$ and the the operator $C$ with domain $\mathcal D(C^{-1})$ given in \cite{ali-antoine-gazeau} are identified here to  $\R^{2}$, and $I_{\mathfrak H}$ the identity operator  on $\mathfrak H = L^{2}(\R)$, respectively, with $\mathcal D(C^{-1}) = \mathfrak H $ and $\mathcal D(C^{-1})^{\dag} = \overline{\mathfrak H}$.
Consider an element in $B_{2}(\mathfrak H)$ of the type $|\phi\rangle \langle \psi|$, with $\phi, \psi\in \mathfrak H$ and let $f = \mathcal W(|\phi\rangle \langle \psi|)$.

For $\phi', \psi' \in \mathfrak H$, one has from the definition of $\mathcal W$ in (\ref{map1}) 
\begin{eqnarray}{\label{map03}}
\int_{\R}\int_{\R}\langle \phi'|U(x,y)\psi' \rangle \mathcal W(|\phi\rangle \langle \psi|)(x,y)dxdy
&=& \int_{\R}\int_{\R}\langle \phi'|U(x,y)\psi' \rangle Tr(U(x,y)^{*}|\phi\rangle \langle \psi|)\cr
&&\times dxdy \cr
&=& \int_{\R}\int_{\R}\overline{\langle \phi|U(x,y)\psi \rangle}\langle \phi'|U(x,y)\psi' \rangle dxdy,
\end{eqnarray}
by the orthogonality relations one gets
\begin{eqnarray}{\label{map3}}
\int_{\R}\int_{\R}\langle \phi'|U(x,y)\psi' \rangle \mathcal W(|\phi\rangle \langle \psi|)(x,y)dxdy  = \langle \phi'|\phi \rangle \langle \psi|\psi' \rangle.
\end{eqnarray}
The relation $|\langle \phi'|\phi \rangle \langle \psi|\psi' \rangle| \leq ||\phi'||||\psi'||||\phi||||\psi||$, implies that
\begin{eqnarray}
\left|\int_{\R}\int_{\R}\langle \phi'|U(x,y)\psi' \rangle \mathcal W(|\phi\rangle \langle \psi|)(x,y)dxdy \right| \leq ||\phi'||||\psi'||||\phi||||\psi||.
\end{eqnarray}
Then, the relation (\ref{map3}) holds for all $\phi',\psi' \in \mathfrak H$. Thus, as a weak integral, one has
\begin{eqnarray}
|\phi\rangle \langle \psi| &=& \int_{\R}\int_{\R}U(x,y)\mathcal W(|\phi\rangle \langle \psi|)(x,y)dxdy \cr
&=& \mathcal W^{-1}f
\end{eqnarray}
Then, the inverse of $\mathcal W$ is defined on the dense set of vectors $f \in L^{2}(\R^{2}, dxdy)$, comprising the image of $\mathfrak H \otimes \overline{\mathfrak H}$, the inverse map $\mathcal W^{-1}$ is such that
\begin{eqnarray}
&&\mathcal W^{-1}:L^{2}(\R^{2}, dxdy) \rightarrow \mathfrak H \otimes \overline{\mathfrak H} \cr
&&\mathcal W^{-1} f =  \int_{\R}\int_{\R}U(x,y)\mathcal W(|\phi\rangle \langle \psi|)(x,y)dxdy.
\end{eqnarray}

\section{Unitary transforms and operator representations}\label{sec3}
This paragraph is devoted to  the
operator representations related to the unitary transforms applied to the quantum Hamiltonians describing the physical system. From the Wigner transform defined on the Hilbert space of Hilbert-Schmidt operators in the context of Tomita-Takesaki modular theory, the related counterparts of both discrete and continuous spectra operators, expressed in the Fock helicity representation,  are determined, respectively. In addition, the Hilbert space realization of the linear operator of the continuous spectrum is also  provided. 
\subsection{Wigner transform and Hamiltonian representations}\label{sec030}

Let $\hat{\mathfrak H} = \mathfrak H\otimes \tilde{\mathfrak  H}$ 
be the Hilbert space with orthonormal basis 
$\{\Phi_{n} \otimes \phi_{\alpha_{k}}\}, $ where $\mathfrak H$ is the Hilbert space with orthonormal basis  $\{\Phi_{n}\}^{\infty}_{n=0}$ and  $\tilde{\mathfrak  H}$ the Hilbert space with orthonormal basis  $\{\phi_{\alpha_{k}}, \alpha_{k} 
\in \R\}, \alpha$ replaced by $\alpha_{k}$, see Eqs. (\ref{es17}) and (\ref{eig003}). Consider $\mathcal B_{2}(\mathfrak H)\simeq \mathfrak H \otimes \bar{\mathfrak H}$ and  $\mathcal B_{2}(\hat{\mathfrak H})\simeq \hat{\mathfrak H} \otimes \bar{\hat{\mathfrak H}}$, the spaces of Hilbert-Schmidt operators on $\mathfrak H$ and $\hat{\mathfrak H}$,  respectively.
Take $U(x, y)$ given in (\ref{unitp001}) and let  $\mathcal Q_1 = \frac{1}{2\sqrt{M\omega_c \hbar}}(b + b^{\dag}), \, \mathcal P_1 = \frac{i}{2\sqrt{M\omega_c\hbar}}(b^{\dag} - b)$ (see Eq. (\ref{commuta000})),  with $[\mathcal  Q_1,  \mathcal P_1] = \imath \mathbb I_{\mathfrak H_{s}}$. Besides, let $\mathcal   Q_2 = \frac{1}{2\sqrt{M\omega_c \hbar}}(\mathfrak  b + \mathfrak b^{\dag}), \, \mathcal P_2 = \frac{i}{2\sqrt{M\omega_c\hbar}}(\mathfrak  b^{\dag} - \mathfrak  b)$ (\ref{eqsop001})),  with $[\mathcal Q_2,  \mathcal P_2] = \imath \mathbb I_{\mathfrak H_{s}}$. 

For the linear parts of the quantum hamiltonians,  let us first consider  
the unitary map $\tilde U(x,y) = e^{-\imath (x\mathfrak Q + y \mathfrak  P)}$ with $[\mathfrak Q, \mathfrak  P] = -\imath \mathbb I_{\tilde {\mathfrak H}}$. Then, consider  
$\mathfrak Q_1 = \frac{1}{2\sqrt{M\omega_c \hbar}}(d + d^{\dag})$ and $\mathfrak  P_1 = \frac{i}{2\sqrt{M\omega_c\hbar}}(d^{\dag} - d)$  (see Eq. (\ref{commuta000})) satisfying  $[\mathfrak Q_1 , \mathfrak  P_1] = -\imath \mathbb I_{\mathfrak H_{s}}$. 

Next, take
$\mathfrak Q_2 = \frac{1}{2\sqrt{M\omega_c \hbar}}(\mathfrak d + \mathfrak d^{\dag})$ and $\mathfrak  P_2 = \frac{i}{2\sqrt{M\omega_c\hbar}}(\mathfrak d^{\dag} - \mathfrak d)$  (see Eq. (\ref{eqsop001})) satisfying  $[\mathfrak Q_2 , \mathfrak  P_2] = -\imath \mathbb I_{\mathfrak H_{s}}$. Then, the linear operators $\hat T_1$ and $\hat T_2$ given in (\ref{tfunc1}) and (\ref{linp001}), respectively, write
\begin{eqnarray}\label{linp003}
 \hat T_1 &=&
\frac{\lambda}{2M}(d^{\dag} + d) + \frac{\lambda^{2}}{2M} = \lambda\sqrt{\frac{\omega_c \hbar}{M}}\mathfrak{Q}_1 + \frac{\lambda^{2}}{2M}, \crcr \hat T_2 &=&
\frac{\lambda}{2M}(\mathfrak d^{\dag} + \mathfrak d) + \frac{\lambda^{2}}{2M} = \lambda\sqrt{\frac{\omega_c \hbar}{M}}\mathfrak{Q}_2 + \frac{\lambda^{2}}{2M}.
\end{eqnarray}
Given any vector $\hat{X} \in \mathcal B_{2}(\hat{\mathfrak H})$, 
$\hat{X} = |\Phi \otimes \phi_{\alpha_{k}}\rangle \langle \Psi \otimes \psi_{\alpha_{k}}|$, we have the map $\hat{\mathcal W}$ on $\mathcal B_{2}(\hat{\mathfrak H})$ using  $\tilde U(x,y)$ given as follows:
\begin{eqnarray}{\label{egal2}}
(\hat{\mathcal W}\hat{X})(x,y) &:=& \frac{1}{(2\pi)}\langle (U(x,y)\otimes \tilde U(x,y)) \Psi \otimes 
\phi_{\alpha_{k}}| \Phi \otimes \psi_{\alpha_{k}}\rangle_{\hat{\mathfrak H}} \cr 
                    &=& \frac{1}{(2\pi)}\langle U(x,y)\Psi|\Phi 
\rangle_{\mathfrak H} \otimes 
\langle 
\tilde U(x,y)) \phi_{\alpha_{k}}|\psi_{\alpha_{k}} \rangle_{\tilde{\mathfrak H}},
\end{eqnarray}
where for the harmonic oscillator part $\hat H_{1_{OSC}}$ (\ref{equah1osc}) of the Hamiltonian $\hat H_1$, from the unitary operator $U(x, y)$, we get 
\begin{equation}{\label{equa28}}
{\mathcal W}Q^{2} \vee I_{{\mathfrak H}} {{\mathcal W}}^{-1} =  \mathcal Q^{2}_{1}, \; \,\,\,
{\mathcal W}P^{2} \vee I_{{\mathfrak H}}{\mathcal W}^{-1} = \mathcal P^{2}_{1}, 
\end{equation}
affording
\begin{equation}{\label{equa30}}
{\mathcal W}\left(\frac{1}{4M}({\hat{Q}}^{2} + {\hat{P}}^{2}) \right)\vee \mathbb{I}_{\mathfrak H}
{\mathcal W}^{-1} = \frac{1}{4M}(\hat{\mathcal{Q}}^{2}_{1} + \hat{\mathcal{P}}^{2}_{1}), \quad [\hat{\mathcal{Q}}_{1},  \hat{\mathcal{P}}_{1}] = 2i M\omega_c \hbar  \mathbb I_{\mathfrak H_{s}}.
\end{equation}
Besides, for the linear parts,  from 
 $\tilde U(x,y) = e^{-\imath (x\mathfrak Q + y \mathfrak  P)}$, (\ref{map1}) and (\ref{linp003}) together, using the map $\tilde{\mathcal W}: \mathcal B_{2}(\tilde\mathfrak H) \rightarrow L^{2}(\R^{2}, dxdy) $ providing analogous expressions as in (\ref{equa28}) for $\mathfrak{Q}_1$,  we have
\begin{eqnarray}{\label{equa32}}
(\tilde{\mathcal W} \hat T_1 \vee I_{\tilde{\mathfrak H}} (|\phi_{\alpha_{k}}\rangle\langle\phi_{\alpha_{k}}|)(x, y) &=&  
\frac{1}{(2\pi)^{1/2}}\langle\tilde U(x,y) \phi_{\alpha_{k}}|\hat T_1\phi_{\alpha_{k}} \rangle_{\tilde{\mathfrak H}} \cr
&=& \mathcal{E}_{\alpha_{k}}\frac{1}{(2\pi)^{1/2}}\langle\tilde U(x,y) \phi_{\alpha_{k}}|\phi_{\alpha_{k}} \rangle_{\tilde{\mathfrak H}},\quad \hat T_1|\phi_{\alpha_{k}}\rangle =  \mathcal{E}_{\alpha_{k}}|\phi_{\alpha_{k}}\rangle\cr
&=&  \mathcal{E}_{\alpha_{k}}\tilde{\mathcal W}(|\phi_{\alpha_{k}}\rangle\langle\phi_{\alpha_{k}}|)(x, y)\cr
&=& \hat T_1\tilde{\mathcal W}|\phi_{\alpha_{k}}\rangle\langle\phi_{\alpha_{k}}|)(x, y).
\end{eqnarray}
Thus, from (\ref{equa30}) and (\ref{equa32}) together, we get
\begin{equation}
\mathcal W\left(\frac{1}{4M}({\hat{Q}}^{2} + {\hat{P}}^{2})\right)\vee \mathbb{I}_{\mathfrak H}
{\mathcal W}^{-1} - \tilde{\mathcal W}\hat{T}\vee\mathbb{I}_{\tilde{\mathfrak H}}\tilde{\mathcal W}^{-1}  = \frac{1}{4M}({\hat{\mathcal Q}}^{2}_{1} + {\hat{\mathcal P}}^{2}_{1}) - \hat{T}_{1},
\end{equation}
i.e.,
\begin{equation}\label{wign050}
\mathcal W\left(\hat{H}_{OSC} \right)\vee \mathbb{I}_{\mathfrak H} \mathcal W^{-1}  - \tilde{\mathcal W}\hat{T}\vee \mathbb{I}_{\tilde{\mathfrak H}}\tilde{\mathcal W}^{-1}
 = \hat{H}_{1_{OSC}} - \hat{T}_{1},
\end{equation}

where $\hat{H}_{1_{OSC}} = \frac{1}{4M}({\hat{\mathcal Q}}^{2}_{1} + {\hat{\mathcal P}}^{2}_{1})$.

In the symmetric gauge $\overrightarrow{A} = \left(-\frac{B}{2}y,\frac{B}{2}x \right)$, we get for the Hamiltonian $\hat H_2$
\begin{equation}\label{wign055}
\mathcal W\mathbb{I}_{\mathfrak H}\vee\left(\hat{H}_{OSC} \right)\mathcal W^{-1}  - \tilde{\mathcal W}\mathbb{I}_{\tilde{\mathfrak H}}\vee\hat{T}\tilde{\mathcal W}^{-1}
 = \hat{H}_{2_{OSC}} - \hat{T}_{2},
\end{equation}
where $\hat{H}_{2_{OSC}} = \frac{1}{4M}({\hat{\mathcal Q}}^{2}_{2} + {\hat{\mathcal P}}^{2}_{2})$.
\brmk
From the Eqs.(\ref{wign050}) and (\ref{wign055}) together, it follows that the set of operators $\{{\hat{\mathcal Q}}_{1},  {\hat{\mathcal P}}_{1}, \mathfrak Q_{1}\}$ and $\{{\hat{\mathcal Q}}_{2},  {\hat{\mathcal P}}_{2}, \mathfrak Q_{2}\}$, see (\ref{linp003}),  generate the von Neumann algebras $\mathfrak A_{+}$ and $\mathfrak A_{-}$, respectively,   with $\mathfrak A_{+} = \hat{\mathcal W}\mathfrak A_{l}\hat{\mathcal W}^{-1}$ and $\mathfrak A_{-} = \hat{\mathcal W}\mathfrak A_{r}\hat{\mathcal W}^{-1}$, where $
\mathfrak A_{l} = \left\{A_{l} = A \vee I| A \in \mathcal L(\mathfrak H)\right\}$, $
\mathfrak A_{r} = \left\{A_{r} = I \vee A| A \in \mathcal L(\mathfrak H)\right\}$, and $\mathcal L(\mathfrak H)$ the set of bounded operators on $\mathfrak H$. In addition, with the transform $\mathcal W: \mathcal B_{2}(\mathfrak H) \rightarrow L^{2}(\R^{2}, dxdy)$,   physically, the two commuting algebras $\mathcal W\mathfrak A_{l}\mathcal W^{-1}$
 and $\mathcal W\mathfrak A_{r}\mathcal W^{-1}$, generated by $\{{\hat{\mathcal Q}}_{1},  {\hat{\mathcal P}}_{1}\}$ and $\{{\hat{\mathcal Q}}_{2},  {\hat{\mathcal P}}_{2}\}$, correspond to the two
directions of the magnetic field given by the symmetric gauges $\overrightarrow{A} = \left(\frac{B}{2}y, -\frac{B}{2}x \right)$ and  $\overrightarrow{A} = \left(-\frac{B}{2}y, \frac{B}{2}x \right)$, respectively \cite{ab1, abh}.
\ermk
\subsection{Helicity quantum Hamiltonians}

Let's define the quadrature operators as follows
\begin{eqnarray}\label{fock000}
 Q_{1} &=& \frac{1}{\sqrt{2}}(b+b^{\dag}), \qquad P_{1} = \frac{i }{\sqrt{2}}(b^{\dag} - b), \qquad [b, b^{\dag}] = 2M\omega_c, \cr
 \tilde Q_{1} &=& \frac{1}{\sqrt{2}}(\mathfrak d+ \mathfrak d^{\dag}), \qquad \tilde P_{1} = \frac{i }{\sqrt{2}}(\mathfrak d^{\dag} - \mathfrak d),\qquad [\mathfrak d, \mathfrak d^{\dag}] = 2M\omega_c, \cr
 Q_{2} &=& \frac{1}{\sqrt{2}}(\mathfrak l+\mathfrak l^{\dag}), \qquad \;  P_{2} = \frac{i }{\sqrt{2}}(\mathfrak l^{\dag} - \mathfrak l), \qquad \; \; [\mathfrak l, \mathfrak l^{\dag}] =  2M\omega_c, \cr
 \tilde Q_{2} &=& \frac{1}{\sqrt{2}}(k+k^{\dag}), \qquad \tilde P_{2} = \frac{i }{\sqrt{2}}(k^{\dag} - k), \qquad [k, k^{\dag}] =  2M\omega_c
\end{eqnarray}
satisfying the commutation relations:  
\begin{equation}\label{fock001}
[Q_{k}, P_{j}] = 2  M  \omega_c i\mathbb I_{\mathfrak H_{s}} \delta_{kj} = [\tilde Q_{k}, \tilde P_{j}], \quad k, j = 1, 2
\end{equation}
have been also made, with $\mathbb I_{\mathfrak H_{s}}$ denoting the identity operator on the Hilbert space $\mathfrak H_{s} = L^2(\mathbb R^2, dxdy)$.

Let us introduce now the following observables $\{Q_{\pm}, P_{\pm}\}$ and $\{\tilde Q_{\pm}, \tilde P_{\pm}\}$ related to the quantum Hamiltonians, with $\hbar = 1$, 
\begin{eqnarray}\label{fock003}\left(
 \begin{array}{c}
 Q_+ \\
P_+ \\
\tilde Q_{+}\\
\tilde P_{+} \\
 \end{array}
\right) = \frac{1}{\sqrt{M\omega_c}}\left(
 \begin{array}{c}
 Q_1 \\
P_1 \\
Q_{2}\\
P_{2} \\
 \end{array}
\right), \qquad \left(
 \begin{array}{c}
 Q_- \\
P_- \\
\tilde Q_{-}\\
\tilde P_{-} \\
 \end{array}
\right) = \frac{1}{\sqrt{M\omega_c}}\left(
 \begin{array}{c}
\tilde  Q_1 \\
\tilde P_1 \\
\tilde Q_{2}\\
\tilde P_{2} \\
 \end{array}
\right), 
\end{eqnarray}
and the   annihilation and creation operators $\{ A_{\pm}, A^{*}_{\pm}\}$ and $\{\tilde A_{\pm}, \tilde A^{*}_{\pm}\}$ acting in the helicity Fock spaces given by
\begin{eqnarray}\label{fock005}
A_{+} &=& \frac{1}{\sqrt{2}}(Q_{+} + i P_{+}), \qquad A^{*}_{+} = \frac{1}{\sqrt{2}}(Q_{+} - i P_{+}),\cr 
A_{-} &=& \frac{1}{\sqrt{2}}(i Q_{-} -  P_{-}), \qquad A^{*}_{-} = \frac{1}{\sqrt{2}}(-i Q_{-} - P_{-}),  \cr
\tilde A_{+} &=& \frac{1}{\sqrt{2}}(\tilde Q_{+} + i \tilde P_{+}), \qquad \tilde A^{*}_{+} = \frac{1}{\sqrt{2}}(\tilde Q_{+} - i \tilde P_{+}) \cr
\tilde A_{-} &=& \frac{1}{\sqrt{2}}(i \tilde Q_{-} - \tilde P_{-}), \qquad
\tilde A^{*}_{-} = \frac{1}{\sqrt{2}}(-i \tilde Q_{-} - \tilde P_{-}) 
\end{eqnarray}
which satisfy
\begin{equation}\label{fock007}
[A_{\pm}, A^{*}_{\pm}] = 2  = [\tilde A_{\pm}, \tilde A^{*}_{\pm}] \quad \mbox{and} \quad [Q_{\pm}, P_{\pm}] = 2 i \mathbb I_{\mathfrak H_{s}} = [\tilde Q_{\pm}, \tilde P_{\pm}] 
\end{equation}
with all other commutators being zero.

In  the scalar potential  $ \Phi(x,y) = Ex$.  We obtain in the symmetric gauges $\overrightarrow{A} = \left(-\frac{B}{2}y, \frac{B}{2}x \right)$ and  $\overrightarrow{A} = \left(\frac{B}{2}y, -\frac{B}{2}x \right)$, from Eqs. (\ref{fock000})-(\ref{fock007}), the following expressions:
\begin{eqnarray}{\label{form01}}
H_{1} &=& \frac{1}{4M}(b^{\dag}b + bb^{\dag}) - \frac{\lambda}{2M}(d^{\dag} + d) - \frac{\lambda^{2}}{2M}, \crcr
 &=&   \frac{1}{4M}(Q^{2}_{1} + P^{2}_{1}) + \frac{\lambda}{M\sqrt{2}}\tilde P_{2} - \frac{\lambda^{2}}{2M}, \qquad d^{\dag} + d = \sqrt{2}\tilde P_{2} = 
\sqrt{2M\omega_{c}} \tilde P_{-}\crcr
&=& \frac{\omega_{c}}{4}(Q^{2}_{+} + P^{2}_{+}) - \frac{\bar \lambda}{2}(\tilde A_{-} + \tilde A^{*}_{-})  - \frac{\lambda^{2}}{2M} := H_{+},
\end{eqnarray}

\begin{eqnarray}{\label{form02}}
\tilde H_{1} &=& \frac{1}{4M}(\mathfrak  d^{\dag}\mathfrak  d + \mathfrak  d \mathfrak  d^{\dag}) - \frac{\lambda}{2M}(\mathfrak  b^{\dag} + \mathfrak  b) - \frac{\lambda^{2}}{2M}, \crcr
&=& \frac{1}{4M}(\tilde Q^{2}_{1}+  \tilde P^{2}_{1}) - \frac{\lambda}{M\sqrt{2}}Q_{2} - \frac{\lambda^{2}}{2M}, \qquad  \mathfrak  b^{\dag} + \mathfrak  b = \sqrt{2}Q_{2}  
= \sqrt{2M\omega_{c}} \tilde Q_{+}\crcr
&=& \frac{\omega_{c}}{4}(Q^{2}_{-} + P^{2}_{-}) - \frac{\bar \lambda}{2}(\tilde A_{+} + \tilde A^{*}_{+})  - \frac{\lambda^{2}}{2M}:= H_{-},
\end{eqnarray}
respectively, where $\bar \lambda = \lambda \sqrt{\frac{\omega_{c}}{M}}$. In  the case of the scalar potential given by $ \Phi(x,y) = Ey$, we obtain in the symmetric gauges $\overrightarrow{A} = \left(-\frac{B}{2}y, \frac{B}{2}x \right)$ and  $\overrightarrow{A} = \left(\frac{B}{2}y, -\frac{B}{2}x \right)$,

\begin{eqnarray}{\label{form03}}
H_{2} &=& \frac{1}{4M}(\mathfrak  l^{\dag} \mathfrak  l +  \mathfrak  l  \mathfrak  l^{\dag}) - \frac{\lambda}{2M}(\mathfrak  d^{\dag} +  \mathfrak  d) - \frac{\lambda^{2}}{2M}, \crcr
&=& \frac{1}{4M}( Q^{2}_{2}+ P^{2}_{2}) - \frac{\lambda}{M\sqrt{2}} \tilde Q_{1} - \frac{\lambda^{2}}{2M}, \quad \mathfrak  d^{\dag} +  \mathfrak  d = \sqrt{2}\tilde Q_{1} 
= \sqrt{2M\omega_{c}} Q_{-}\crcr
&=& \frac{\omega_{c}}{4}(\tilde Q^{2}_{+} + \tilde P^{2}_{+}) - \frac{i \bar \lambda}{2}(A^{*}_{-} - A_{-})  - \frac{\lambda^{2}}{2M}:= \tilde H_{+}.
\end{eqnarray}

\begin{eqnarray}{\label{form04}}
\tilde H_{2} &=& \frac{1}{4M}(k^{\dag} k +  k  k^{\dag}) - \frac{\lambda}{2M}( l^{\dag} +  l) - \frac{\lambda^{2}}{2M}, \crcr
&=& \frac{1}{4M}(\tilde Q^{2}_{2}+  \tilde P^{2}_{2}) - \frac{\lambda}{M\sqrt{2}} P_{1} - \frac{\lambda^{2}}{2M}, \qquad l^{\dag}+ l = \sqrt{2}P_{1}
 = \sqrt{2M\omega_{c}} P_{+}\crcr
&=& \frac{\omega_{c}}{4}(\tilde Q^{2}_{-} + \tilde P^{2}_{-}) - \frac{i \bar \lambda}{2}(A^{*}_{+} - A_{+})  - \frac{\lambda^{2}}{2M}:= \tilde H_{-},
\end{eqnarray}
respectively. 

From the results of the paragraph \ref{sec030}, the pairs of Hamiltonians $\{H_+, \tilde{H}_+\}$ and $\{H_-, \tilde{H}_-\}$, expressed from (\ref{form01})-(\ref{form04}), corresponding to the  magnetic field directed along the positive and  negative $z$-direction, respectively, satisfy analogous relations as in (\ref{wign050}) and (\ref{wign055}).

\subsection{Another representation and continuous spectra Hilbert spaces}
The position and momentum operators $P = \sqrt{2}\mathcal P$ and $Q = \sqrt{2}\mathcal Q$ such that  $ [\mathcal Q,  \mathcal P] = \imath \mathbb I_{\mathfrak H}$, where  we fix $\{P_{\pm}, Q_{\pm}\} \equiv \{P, Q\} \equiv \{\tilde P_{\pm}, \tilde Q_{\pm}\}$  with $\{P_{\pm}, Q_{\pm}\}$ and $\{\tilde P_{\pm}, \tilde Q_{\pm}\}$ satisfying the commutation relations (\ref{fock007}),  in the Schr\"{o}dinger representation given on $\mathfrak H = L^{2}(\R)$ are such that one gets, from the Wigner transform, the following relations:
\begin{eqnarray}{\label{wign030}}
\mathcal W \left(
\begin{array}{c}
I_{\mathfrak H} \vee Q\\
I_{\mathfrak H} \vee P
\end{array}
\right)\mathcal W^{-1} = \left(
\begin{array}{c}
P_{+} \\
Q_{+}
\end{array}
\right), \;\; \mathcal W \left(
\begin{array}{c}
Q \vee I_{\mathfrak H} \\
P \vee I_{\mathfrak H}
\end{array}
\right)\mathcal W^{-1} = \left(
\begin{array}{c}
Q_{-} \\
P_{-}
\end{array}
\right), \nonumber \\
\end{eqnarray}
\begin{eqnarray}{\label{wign032}}
\mathcal W \left(
\begin{array}{c}
I_{\mathfrak H} \vee Q\\
I_{\mathfrak H} \vee P
\end{array}
\right)\mathcal W^{-1} = \left(
\begin{array}{c}
\tilde P_{+} \\
\tilde Q_{+}
\end{array}
\right), \;\; \mathcal W \left(
\begin{array}{c}
Q \vee I_{\mathfrak H} \\
P \vee I_{\mathfrak H}
\end{array}
\right)\mathcal W^{-1} = \left(
\begin{array}{c}
\tilde Q_{-} \\
\tilde P_{-}
\end{array}
\right).\nonumber \\
\end{eqnarray}

Consider the map
\begin{eqnarray}
\mathcal U:L^{2}(\R^{2}, dxdy) \rightarrow \mathfrak H \otimes \mathfrak H = L^{2}(\R) \otimes L^{2}(\R),
\end{eqnarray}
with $\mathcal U = \mathcal I \circ \mathcal W^{-1}$, where $\mathcal I:\mathfrak H \otimes \overline{\mathfrak H} \rightarrow \mathfrak H \otimes \mathfrak H$, such that for a given vector $|\phi\rangle \langle \psi| \in \mathfrak H \otimes \overline{\mathfrak H}$, $\mathcal I(\phi(x)\overline{\psi(y)}) = \phi(x)\psi(y), \; x,y \in \R, \phi, \psi \in \mathfrak H$. Then, for $X = |\phi\rangle \langle \psi| \in \mathfrak H \otimes \overline{\mathfrak H}$, one has
\begin{eqnarray}
\mathcal U(\mathcal W(Q \otimes I_{\mathfrak H}(X))(x,y) &=& \mathcal U(\mathcal W(Q \otimes I_{\mathfrak H}(|\phi\rangle \langle \psi|))(x,y) \cr
&=& (Q \otimes I_{\mathfrak H})\phi(x)\psi(y).
\end{eqnarray}
It comes that the Hamiltonians $H_{+}$ and $H_{-}$ given in (\ref{form01}) and (\ref{form02}), from the Eqs. (\ref{wign030}) and (\ref{wign032}) together,  become
\begin{eqnarray}{\label{form08}}
\mathcal U H_{+} \mathcal U^{-1} &=& \frac{\omega_{c}}{2}\mathcal U(A^{*}_{+}A_{+} + 1)\mathcal U^{-1} -  \mathcal U\left[\frac{\bar \lambda}{2}(\tilde A_{-} + \tilde A^{*}_{-})  + \frac{\lambda^{2}}{2M}\right] \mathcal U^{-1}\crcr
&=&  \frac{\omega_{c}}{2}(\mathcal  N_{+} + 1) \otimes I_{\mathfrak H_C} -  I_{\mathfrak H_D}\otimes \left(\frac{\bar \lambda}{2}(\tilde c + \tilde c^{\dag}) +  \frac{\lambda^{2}}{2M}\right)   \crcr
&=&  H_{1_{D}} \otimes I_{\mathfrak H_C} +  I_{\mathfrak H_D} \otimes H_{1_{C}}, 
\end{eqnarray}
\begin{eqnarray}{\label{form09}}
\mathcal U H_{-} \mathcal U^{-1} &=& \frac{\omega_{c}}{2} \mathcal U(A^{*}_{-}A_{-} + 1) \mathcal U^{-1}  - \mathcal U\left[\frac{\bar \lambda}{2}(\tilde A_{+} + \tilde A^{*}_{+})  + \frac{\lambda^{2}}{2M}\right]\mathcal U^{-1} \crcr
&=& \frac{\omega_{c}}{2}(\mathcal  N_{-} + 1) \otimes I_{\mathfrak H_C}  - I_{\mathfrak H_D} \otimes \left(\frac{\bar \lambda}{2}(\tilde a + \tilde a^{\dag}) + \frac{\lambda^{2}}{2M}\right)\crcr
&=& H_{2_{D}} \otimes I_{\mathfrak H_C} + I_{\mathfrak H_D} \otimes H_{2_{C}} ,
\end{eqnarray}
where $\mathcal N_{+} = a^{\dag}a$ and $\mathcal N_{-} = c^{\dag}c$, respectively.

For the operators $\tilde A_{\pm}, \tilde A^{*}_{\pm}$, one obtains  in the gauge ${\bf A} = \left(-\frac{B}{2}y, \frac{B}{2}x \right)$ and   ${\bf A} = \left(\frac{B}{2}y, -\frac{B}{2}x \right)$, the  following relations
\begin{equation}
\mathcal U \tilde A_{+} \mathcal U^{-1} : = I_{\mathfrak H_D} \otimes \tilde a \qquad
\mathcal U \tilde A^{*}_{+} \mathcal U^{-1} : = I_{\mathfrak H_D} \otimes \tilde a^{\dag}, 
\end{equation}
and
\begin{equation}
\mathcal U \tilde A_{-} \mathcal U^{-1}  : = I_{\mathfrak H_D} \otimes  \tilde c \qquad
\mathcal U A^{*}_{-} \mathcal U^{-1} : = I_{\mathfrak H_D} \otimes \tilde c^{\dag}, 
\end{equation}
respectively.

Thus,  the Hamiltonians $\tilde H_{+}$ and $\tilde H_{-}$, given in (\ref{form03}) and (\ref{form04}), become
\begin{eqnarray}{\label{form010}}
\mathcal U \tilde H_{+} \mathcal U^{-1} &=& \frac{\omega_{c}}{2}\mathcal U(\tilde A^{*}_{+}\tilde A_{+} + 1)\mathcal U^{-1} -  \mathcal U\left[i \frac{\bar \lambda}{2}(A_{-} - A^{*}_{-})  + \frac{\lambda^{2}}{2M}\right] \mathcal U^{-1}\crcr
&=&  \frac{\omega_{c}}{2} (\tilde{\mathcal  N}_{+} + 1)\otimes I_{\mathfrak H_C}  - I_{\mathfrak H_D} \otimes\left(i \frac{\bar \lambda}{2}(c - c^{\dag}) + \frac{\lambda^{2}}{2M}\right) \crcr
&=&  \tilde H_{1_{D}} \otimes I_{\mathfrak H_C} +   I_{\mathfrak H_D} \otimes \tilde H_{1_{C}}, 
\end{eqnarray}
\begin{eqnarray}{\label{form011}}
\mathcal U \tilde H_{-} \mathcal U^{-1} &=& \frac{\omega_{c}}{2} \mathcal U(\tilde A^{*}_{-}\tilde A_{-} + 1) \mathcal U^{-1}  - \mathcal U\left[\frac{\bar \lambda}{2}(A_{+} -  A^{*}_{+})  + \frac{\lambda^{2}}{2M}\right]\mathcal U^{-1} \crcr
&=& \frac{\omega_{c}}{2}(\tilde{\mathcal  N}_{-} + 1)\otimes I_{\mathfrak H_C} - I_{\mathfrak H_D} \otimes \left(\frac{\bar \lambda}{2}(a -  a^{\dag}) + \frac{\lambda^{2}}{2M}\right)\crcr
&=& \tilde H_{2_{D}} \otimes I_{\mathfrak H_C} + I_{\mathfrak H_D} \otimes \tilde H_{2_{C}} ,
\end{eqnarray}
where $\tilde{\mathcal N}_{+} = \tilde a^{\dag}\tilde a$ and $\tilde{\mathcal N}_{-} = \tilde c^{\dag}\tilde c$, respectively, with
\begin{equation}
\mathcal U  A_{+} \mathcal U^{-1} : = I_{\mathfrak H_D} \otimes  a \qquad
\mathcal U  A^{*}_{+} \mathcal U^{-1} : = I_{\mathfrak H_D} \otimes  a^{\dag}, 
\end{equation}
and
\begin{equation}
\mathcal U  A_{-} \mathcal U^{-1}  : = I_{\mathfrak H_D} \otimes c \qquad
\mathcal U A^{*}_{-} \mathcal U^{-1} : = I_{\mathfrak H_D} \otimes c^{\dag}.
\end{equation}
Introducing the position operators $\hat{q_{1}} = \frac{1}{\sqrt{2}}(\tilde c^{\dag} + \tilde c)$ and $\hat{q_{2}} = \frac{1}{\sqrt{2}}(\tilde a + \tilde a^{\dag})$, one gets
\begin{eqnarray}
H_{1_{C}} &=& - \frac{\bar \lambda}{2}(\tilde c^{\dag} + \tilde c) - \frac{\lambda^{2}}{2} = - \frac{\bar \lambda}{\sqrt{2}}\hat{q_{1}} - \frac{\lambda^{2}}{2}\cr
H_{2_{C}} &=& - \frac{\bar \lambda}{2}(\tilde a + \tilde a^{\dag}) - \frac{\lambda^{2}}{2} = -  \frac{\bar \lambda}{\sqrt{2}}\hat{q_{2}} - \frac{\lambda^{2}}{2}.
\end{eqnarray}
The basis of eigenvectors of $\hat{q_{1}}$ and $\hat{q_{2}}$  are given by
\begin{eqnarray}{\label{repr06}}
\mathfrak H_{C} = L^{2}(\R) &=& \left\{\phi(\alpha)\, / \, ||\phi||^{2} = \int_{\R}d\alpha |\phi(\alpha)|^{2} < \infty \right\}, \crcr
\phi(\alpha) &=& \langle \alpha|\phi \rangle, \quad \hat{q_{i}}|\alpha\rangle = \alpha |\alpha\rangle, \; i = 1,2, \qquad
\langle \alpha'|\alpha \rangle = \delta(\alpha' - \alpha).
\end{eqnarray}

In the sequel, from the properties of the  map $\mathcal W$, the Hilbert space representation $\mathfrak  H_{s} = L^2(\R^2, dxdy)$ spanned by the discrete spectrum eigenvectors $|\Psi_{nl}\rangle$ is equivalent through $\mathcal U$ to the Hilbert space $L^2(\R) \otimes L^2(\R)$, where $\mathfrak H  = L^2(\R) = span\{|\Phi_n\rangle  = |n\rangle\}_{n=0}^{\infty}$, 
see (\ref{es17}). 
{Therefore, the Hilbert space $\mathfrak  H_D \otimes \mathfrak H_{C} = L^2(\R) \otimes L^2(\R) \otimes L^2(\R)$, with $\mathfrak H_{C}$ given in  (\ref{repr06}),  realizes a  basis  representation for constructing CSs.

\section{Coherent states construction}\label{sec4}
In this section, CSs are constructed on the equivalent Hilbert spaces obtained through unitary transforms in Section \ref{sec3}, by considering the two possible orientations of the magnetic field as in  \cite{aremua-gouba1, aremua-gouba2}.
 
These CSs obey the Gazeau-Klauder criteria, which are full investigated, \cite{gazeau-klauder} that a family of CSs must satisfy.
Besides, using unitary operators, generated by exponentials of  pairs of annihilation and creation operators for the discrete spectra, realizing unitary
irreducible representations of the Weyl-Heisenberg group, combined to infinitesimal displacements operators for the continuous spectra,  GK-CSs are also achieved.

\subsection{Coherent states of Gazeau-Klauder type}
 We consider in this section the  case of the Hamiltonian $\left({H}_{1_{D}} -  \frac{\omega_{c}}{2}  I_{\mathfrak H_{D}} \right) \otimes I_{\mathfrak H_{C}} + I_{\mathfrak H_{D}}\otimes
\left(H_{1_{C}} + \frac{\lambda^{2}}{2} I_{\mathfrak H_{C}}\right)$.

The shifted Hamiltonian $\left({H}_{1_{D}} -  \frac{\omega_{c}}{2}  I_{\mathfrak H_{D}} \right) \otimes I_{\mathfrak H_{C}} + I_{\mathfrak H_{D}}\otimes
\left(H_{1_{C}} + \frac{\lambda^{2}}{2} I_{\mathfrak H_{C}}\right):= \mathcal H_{1_{D}} + \mathcal H_{1_{C}}$ issued from (\ref{form08}) possesses a discrete spectrum given by
$\{E_{n} =
\omega_{c} n, n = 0, 1, 2, \dots\}$ and a continuous spectrum given by $\{ E_{\alpha} = -\frac{\bar \lambda}{\sqrt{2}}\alpha, \alpha \in \R\}$, {where the discrete spectrum Hilbert space is $\mathfrak H_D = span\{|n\rangle\otimes |l\rangle = |n,l\rangle, n, l=0,1,2,\dots\}$} and for the continuous spectrum  $\mathfrak H_{C}= L^{2}(\R) = span\{\phi(\alpha) , \; \alpha \in \R \}$. The  eigenenergies of the shifted Hamiltonian then write
\begin{equation}
\mathcal E_{n, \alpha}  = E_n + E_{\alpha} = \omega_{c} n - \frac{\bar \lambda}{\sqrt{2}}\alpha
\end{equation}
where the condition $\mathcal E_{n, \alpha} \geq 0$ implying $0 \leq \alpha \leq  \frac{\sqrt{2}}{\bar \lambda}\omega_{c} $ is satisfied. 

{The GK-CSs, denoted $|J,\gamma; J', \gamma';l; K_{1}, \theta_{1}\rangle$ as in \cite{aremua-gouba1},  
related to the shifted Hamiltonian with $l = 0,1,2,\dots$ counting the degeneracy of the Landau levels,   
are defined on the Hilbert space $\mathfrak{H}_D \otimes \mathfrak{H}_C$ by
\begin{eqnarray}{\label{vcs00}}
|J,\gamma; J', \gamma';l; K_{1}, \theta_{1}\rangle 
&=& |J,\gamma; J', \gamma';l\rangle \otimes |K_{1}, \theta_{1}\rangle  \cr
&=& \left[\mathcal N(J) \mathcal N(J')\right]^{-1}J'^{l/2}e^{i l \gamma'}\sum^{\infty}_{n=0}\frac{J^{n/2}e^{-i n \gamma}}{\sqrt{n !l !}}|n,l\rangle \cr
&&\otimes  (N(K_{1}))^{-1}\int_{0}^{\infty}\frac{K_{1}^{E_{\alpha_{1}}}}{e^{\frac{1}{2}\beta E^{2}_{\alpha_{1}}}}e^{-i \theta_{1} E_{\alpha_{1}}}|\alpha_{1}\rangle dE_{\alpha_{1}},
\end{eqnarray}
}
with the labelling parameters  chosen such that $0\leq J, J', K_{1} \leq \infty$, $0 \leq \gamma, \gamma' \leq 2\pi$ and $ -\infty <\theta_{1} < \infty$  and the parameter $\beta$ is taken such that $\beta > 0$.
\bpro\label{propcs000}
The normalization constants are given by
\begin{eqnarray} (\mathcal N(J))^2 = \sum^{\infty}_{n=0}\frac{J^{n}}{n !} = e^{J},\qquad (\mathcal N(J'))^2 = \sum^{\infty}_{l=0}\frac{J'^{l}}{l !} = e^{J'}
\end{eqnarray}
and
\begin{eqnarray} (N(K_{1}))^{2} = \int_{0}^{\infty}
e^{2E_{\alpha_{1}} \ln K_{1} - \beta E^{2}_{\alpha_{1}}}dE_{\alpha_{1}} = \frac{1}{2}\sqrt{\frac{\pi}{\beta}}e^{\frac{(\ln K_{1})^{2}}{\beta}}\left[1 - erf\left(\frac{|\ln K_{1}|}{\sqrt{\beta}}\right)\right].\nonumber
\\
\end{eqnarray}
\epro
{\bf Proof.}
Indeed, we have by definition, 
\begin{equation}
\sum^{\infty}_{l=0}\langle J,\gamma;J',\gamma';l|J,\gamma;J',\gamma';l\rangle  = \frac{1}{\mathcal N(J)}\sum^{\infty}_{n=0}\frac{J^{n}}{n !}\frac{1}{\mathcal N(J')}
\sum^{\infty}_{l=0}\frac{J'^{l}}{l !} = 1,
\end{equation}
and 
\begin{eqnarray}
&&\langle K_{1},\theta_{1}|K_{1},\theta_{1}\rangle = 1 \cr
&&\Rightarrow (N(K_{1}))^{-2}\int_{0}^{\infty}\int_{0}^{\infty}
\frac{K_{1}^{E_{\alpha_{1}} + E_{\alpha'_{1}}}}{e^{\frac{1}{2}\beta (E^{2}_{\alpha_{1}} + E^{2}_{\alpha'_{1}})}}e^{-i \theta_{1} (E_{\alpha_{1}} - E_{\alpha'_{1}})}\langle \alpha'_{1}|\alpha_{1}\rangle dE_{\alpha'_{1}}dE_{\alpha_{1}} = 1 \cr
&&\Rightarrow (N(K_{1}))^{-2}\int_{0}^{\infty}\int_{0}^{\infty}
\frac{K_{1}^{E_{\alpha_{1}} + E_{\alpha'_{1}}}}{e^{\frac{1}{2}\beta (E^{2}_{\alpha_{1}} + E^{2}_{\alpha'_{1}})}}e^{-i \theta_{1} (E_{\alpha_{1}} - E_{\alpha'_{1}})}\delta(\alpha'_{1} - \alpha_{1}) dE_{\alpha'_{1}}dE_{\alpha_{1}} = 1 \cr
\cr
&&\Rightarrow (N(K_{1}))^{-2}\int_{0}^{\infty}
e^{2E_{\alpha_{1}} \ln K_{1} - \beta E^{2}_{\alpha_{1}}}dE_{\alpha_{1}} = 1\cr
\cr
&&\Rightarrow (N(K_{1}))^{2} = \int_{0}^{\infty}
e^{2E_{\alpha_{1}} \ln K_{1} - \beta E^{2}_{\alpha_{1}}}dE_{\alpha_{1}} = \frac{1}{2}\sqrt{\frac{\pi}{\beta}}e^{\frac{(\ln K_{1})^{2}}{\beta}}\left[1 - erf\left(\frac{|\ln K_{1}|}{\sqrt{\beta}}\right)\right].\nonumber
\\
\end{eqnarray}

$\hfill{\square}$

Next, let us verify that the constructed GK-CSs satisfy all
Klauder's minimal requirements \cite{gazeau-klauder}: (a) continuity in the labelling;
(b) resolution of unity; (c) temporal stability and (d) action identity.
{
\bpro\label{propcs001}
The continuity of the combined CSs follows from the  continuity of the separate states, which are   assumed.  Indeed, from the definition, we have
\begin{equation}
 \| |J,\gamma; J', \gamma';l; K_{1}, \theta_{1}\rangle - |\tilde J,\tilde \gamma; \tilde J', \tilde \gamma';l; \tilde K_{1}, \tilde \theta_{1}\rangle  \|^2
\end{equation}
such that 
\begin{equation}
\lim_{(J,\gamma; J', \gamma';l; K_{1}, \theta_{1})
\rightarrow (\tilde J,\tilde \gamma; \tilde J', \tilde \gamma';l; \tilde K_{1}, \tilde \theta_{1})}\| |J,\gamma; J', \gamma';l; K_{1}, \theta_{1}\rangle - |\tilde J,\tilde \gamma; \tilde J', \tilde \gamma';l; \tilde K_{1}, \tilde \theta_{1}\rangle \|^2
=0.
\end{equation}
\epro
{\bf Proof.} See in the Appendix.
$\hfill{\square}$

}
{
\bpro\label{propcs007}
The GK-CSs (\ref{vcs00}) satisfy on $\mathfrak H_{D} \otimes \mathfrak H_{C} $
the resolution of the identity
\begin{eqnarray}{\label{res00}}
&&\int^{\infty}_{-\infty}
\int^{\infty}_{0}\int^{\infty}_{0}
\int^{\infty}_{0}\int_{0}^{2\pi}\int_{0}^{2\pi}
|J,\gamma; J', \gamma';l; K_{1}, \theta_{1}\rangle \langle J,\gamma; J', \gamma';l; K_{1}, \theta_{1}| \cr
&& \frac{d\gamma}{2\pi}\frac{d\gamma'}{2\pi}\frac{d\theta_{1}}{2\pi}\mathcal N(J)^{2}
\mathcal N(J')^{2}
 N(K_{1})^{2}d\nu(J)d\nu(J')d\rho(K_{1}) = I_{\mathfrak H^l_D}\otimes I_{\mathfrak H_C}
\end{eqnarray}
with $d\rho(K_{1}) = \sigma(K_{1})dK_{1}$, where $\sigma(K_{1})$ is a non-negative weight function $\sigma(K_{1}) \geq 0$ such that
\begin{eqnarray}
\int_{0}^{\infty}K_{1}^{2E_{\alpha_{1}}} \sigma(K_{1}) dK_{1} = e^{\beta E^{2}_{\alpha_{1}}}.
\end{eqnarray}
$I_{\mathfrak H^l_D}, I_{\mathfrak H^n_D}$  are the identity operators on the subspaces $\mathfrak H^n_D, \mathfrak H^l_D$ of $\mathfrak{H}_D$ such that 
\begin{equation}\label{identsubs}
\sum_{n=0}^{\infty}|n,l\rangle \langle n,l| = I_{\mathfrak H^l_D}, \quad \sum_{l=0}^{\infty}|n,l\rangle \langle n,l| = I_{\mathfrak H^n_D}.
\end{equation}
The identity operator $I_{\mathfrak H^l_D}\otimes I_{\mathfrak H_C}$ is the tensor product of
the identity operators $I_{\mathfrak H^l_D}$ and $I_{\mathfrak H_C}$ which act
on the spaces $\mathfrak H^l_D$ and ${\mathfrak H}_{C}$, respectively, corresponding to discrete  and continuous spectra.
\epro
}

{\bf Proof.} See in the Appendix. 
$\hfill{\square}$

{
\bpro\label{propcs003}
The property of temporal stability is given by the relation
\begin{eqnarray}
e^{-i \mathcal H_{1} t}|J,\gamma; J', \gamma';l; K_{1}, \theta_{1}\rangle  &=& e^{-i \mathcal H_{1_{D}}t}|J,\gamma; J', \gamma';l\rangle \otimes e^{-i \mathcal H_{1_{C}}t}|K_{1}, \theta_{1}\rangle  \cr
&=& |J,\gamma + \omega_{c}t;J', \gamma';l;K_{1}, \theta_{1} + t\rangle.
\end{eqnarray}
\epro

{\bf Proof.} See in the Appendix. 
$\hfill{\square}$

}
\bpro\label{propcs005}
The action identity property satisfied by the GK-CSs $|J,\gamma; J', \gamma';l; K_{1}, \theta_{1}\rangle$
is obtained for both Hamiltonians as follows 
\begin{eqnarray}
\sum^{\infty}_{l=0}\langle J,\gamma; J', \gamma';l; K_{1}, \theta_{1}|{\mathcal H}_{1_{D}}|J,\gamma; J', \gamma';l; K_{1}, \theta_{1}\rangle &=&  \omega_c J,\cr
\langle J,\gamma; J', \gamma';l; K_{1}, \theta_{1}| { {\mathcal H}}_{1_{C}}| J,\gamma; J', \gamma';l; K_{1}, \theta_{1}\rangle &=&  {{\mathcal H}}_{1_{C}}(K_{1}). 
\end{eqnarray}
\epro
{\bf Proof.}
See in the Appendix. 
$\hfill{\square}$

Thereby, the CSs $|J,\gamma; J', \gamma';l; K_{1}, \theta_{1}\rangle$ satisfy all Klauder's minimal requirements.

\subsection{Coherent states from  unitary operators}

In this section, we deal with the CSs construction using unitary 
operators defined  for the discrete and continuous spectra, respectively. 

The unitary operators of infinitesimal displacements $D_{\varepsilon}$ and $D_{\eta}$  acting  on $\{|x\rangle \}$ and $\{|y\rangle \}$ representations are given by
\begin{eqnarray}{\label{equa48}}
D_{\varepsilon} = e^{-i \varepsilon(-i \partial_{x})} = e^{- \varepsilon\partial_{x}}, \qquad 
D_{\eta} = e^{-i \eta(-i \partial_{y})} = e^{- \eta\partial_{y}},
\end{eqnarray}
respectively.
In the $\{|x\rangle\}$ representation, one has
\begin{eqnarray}
D_{\varepsilon}\langle x|\alpha \rangle = \langle x|D_{\varepsilon}|\alpha 
\rangle = \langle x-\varepsilon|\alpha \rangle. 
\end{eqnarray}
Replacing $x$ by $x+\varepsilon$, we get
\begin{eqnarray}
D_{\varepsilon}\langle x + \varepsilon|\alpha \rangle = \langle x + \varepsilon|D_{\varepsilon}
|\alpha \rangle 
= \langle x +\varepsilon -\varepsilon|\alpha \rangle  = \langle x|\alpha \rangle := \phi_{\alpha}(x).
\end{eqnarray}
Thereby
\begin{eqnarray}
D_{\varepsilon}\phi_{\alpha}(x + \varepsilon) = \phi_{\alpha}(x).
\end{eqnarray}
In the $\{|y\rangle\}$ representation, we obtain
\begin{eqnarray}
D_{\eta}\phi_{\alpha}(y + \eta) = \phi_{\alpha}(y).
\end{eqnarray}
Define, for the Hamiltonian $\left(\hat {H}_{1_{osc}} -  \frac{\hbar \omega_{c}}{2}  I_{\mathfrak H_{D}} \right)- 
\left(\hat T_{1} - \frac{\lambda^{2}}{2M} I_{\mathfrak H_{C}}\right)$, obtained from (\ref{equah1osc}) and (\ref{tfunc1}), the unitary operator
\begin{eqnarray}
\hat{U}_{1}(z) = U_{1}(z)  \oplus  D_{\varepsilon}
\end{eqnarray}
where  $D_{\varepsilon}$ is given by (\ref{equa48}), and  the unitary operator $U_{1}(z)$ defined on the  Hilbert space $\mathfrak H_{s} = L^{2}(\R^{2},dxdy)$ such that
\begin{eqnarray}
U_{1}(z) =:U_{1}(x,y) =  e^{zb'^{\dag} - \bar{z}b'} = e^{-\frac{1}{2}|z|^{2}}e^{zb'^{\dag}} e^{- \bar{z}b'}, \quad [b',  b'^{\dag}] = \mathbb I_{\mathfrak H_s}, 
\end{eqnarray}
with the annihilation and creation operators $b'$ and  $b'^{\dag}$, see (\ref{es7}) and (\ref{commuta000}),  given by $\displaystyle b' = \frac{b}{\sqrt{2 M\omega_c \hbar}}$ and $\displaystyle b'^{\dag} = \frac{b^{\dag}}{\sqrt{2 
M\omega_c \hbar}}, $
respectively.
\bpro\label{unitopcss000}
Then,  setting $z = J^{1/2}e^{-\imath \gamma}$ and $z' = J'^{1/2}e^{-\imath \gamma'}$ in the definition of the CSs (see \cite{aremua-gouba1}\cite{aremua-gouba2})
\begin{eqnarray}
|J,\gamma;J',\gamma';l;K,\theta;\beta\rangle 
&=& f(K,\theta)|J,\gamma;J',\gamma';l\rangle  + e^{-\imath \beta}g(J, \gamma,J',\gamma') |K, \theta\rangle \cr
&=& f(K,\theta)\left[\mathcal N(J) \mathcal N(J')\right]^{-1/2}J'^{l/2}e^{\imath l \gamma'}\sum^{\infty}_{n=0}\frac{J^{n/2}e^{-\imath n \gamma}}{\sqrt{n !l !}}|\Psi_{nl}\rangle  \cr
&& + e^{-\imath \beta}g(J,\gamma,J',\gamma')\mathcal N_{\rho}(K)^{-1/2}\int^{\infty}_{0}\frac{K^{\epsilon^{-}_{\alpha}/2}e^{\imath \epsilon_{\alpha}\theta}}{\sqrt{\rho(\epsilon^{-}_{\alpha})}} |\epsilon^{-}_{\alpha}\rangle d\epsilon^{-}_{\alpha},
\end{eqnarray}
constructed for the shifted Hamiltonian $\left(\hat {H}_{1_{osc}} -  \frac{\hbar \omega_{c}}{2}  I_{\mathfrak H_{D}} \right)- 
\left(\hat T_{1} - \frac{\lambda^{2}}{2M} I_{\mathfrak H_{C}}\right)$, denoted here as $|z,\bar{z'};l;K,\theta;\beta\rangle$, leads to
\begin{eqnarray}
|z,\bar{z'};l;K,\theta;\beta\rangle  
&=& f(K,\theta)e^{-\frac{|z'|^{2}}{2}}\frac{\bar{z'}^{l}}{\sqrt{l !}}
U_{1}(z)|\Psi_{0l}\rangle  \, \cr
\cr
&+&e^{-i \beta}g(z,\bar{z'})\mathcal N_{\rho}(K)^{-1/2}\int^{\infty}_{0}\frac{K^{\epsilon^{-}_{\alpha}/2}
e^{i \epsilon_{\alpha}\theta}}{\sqrt{\rho(\epsilon^{-}_{\alpha})}}d\epsilon^{-}_{\alpha} \int |x\rangle dx D_{\varepsilon}\phi_{\alpha}(x+\varepsilon). \nonumber
\\
\end{eqnarray}
\epro

{\bf Proof.} See in the Appendix.
$\hfill{\square}$

In the case of the Hamiltonian $\left(\hat {H}_{2_{osc}} -  \frac{\hbar \omega_{c}}{2}  I_{\mathfrak H_{D}} \right)- 
\left(\hat T_{2} - \frac{\lambda^{2}}{2M} I_{\mathfrak H_{C}}\right)$, obtained from (\ref{equah2osc}) and (\ref{linp001}),  the unitary operator is given by
\begin{eqnarray}
\hat{U}_{2}(z') = U_{2}(z')  \oplus   D_{\eta},
\end{eqnarray}
where  $D_{\eta}$ is defined by (\ref{equa48}), and the unitary 
operator $U_{2}(z)$ given on the Hilbert space $\mathfrak H_{s} = L^{2}(\R^{2},dxdy)$ as
\begin{eqnarray}
U_{2}(z') =:U_{2}(x',y') =  e^{\bar{z'}\mathfrak b'^{\dag} - z'\mathfrak b'} = e^{-\frac{1}{2}|z'|^{2}}e^{\bar{z'}\mathfrak b'^{\dag}} 
e^{- z'\mathfrak b'}, \quad [\mathfrak b', \mathfrak b'^{\dag}] = \mathbb I_{\mathfrak H_s}, 
\end{eqnarray}
where $\mathfrak b'$ and  $\mathfrak b'^{\dag}$, see (\ref{eqsop001}),  are  given by $\displaystyle \mathfrak b' = \frac{\mathfrak b}{\sqrt{2 M\omega_c \hbar}}$ and $\displaystyle \mathfrak b'^{\dag} = \frac{\mathfrak b^{\dag}}{\sqrt{2 
M\omega_c \hbar}}, $
respectively. Then, we also have the following result: 
\bpro
The CSs denoted $|z,\bar{z};n;K,\theta;\beta\rangle$ associated to the shifted  Hamiltonian $\left(\hat {H}_{2_{osc}} -  \frac{\hbar \omega_{c}}{2}  I_{\mathfrak H_{D}} \right)- 
\left(\hat T_{2} - \frac{\lambda^{2}}{2M} I_{\mathfrak H_{C}}\right)$,  are obtained  on the $\{|y\rangle\}$ representation as follows:
\begin{eqnarray}
|z,\bar{z'};n;K,\theta;\beta\rangle  &=& f(K,\theta)e^{-\frac{|z|^{2}}{2}}\frac{z^{n}}{\sqrt{n !}}U_{2}(z')
|\Psi_{n0}\rangle \, \cr
\cr
&+&e^{-i \beta}g(z,\bar{z'})\mathcal N_{\rho}(K)^{-1/2}\int^{\infty}_{0}\frac{K^{\epsilon_{\alpha}/2}
e^{i \epsilon_{\alpha}\theta}}{\sqrt{\rho(\epsilon_{\alpha})}}d\epsilon_{\alpha}   
\int |y\rangle dy D_{\eta}\phi_{\alpha}(y + \eta). \nonumber
\\
\end{eqnarray}
\epro

{\bf Proof.} See that of Proposition \ref{unitopcss000}.
$\hfill{\square}$

\bpro
In the case of the Hamiltonian $\hat H_{1_{OSC}} - \hat H_{2_{OSC}}$, we have, using the two unitary operators $U_{1}(z)$ and 
$U_{2}(z)$, on the separable Hilbert space $\hat{\mathfrak H} = \mathfrak H_{s} \otimes \mathfrak H'$, the related CSs given by 
\begin{eqnarray}\label{bcs000}
|z,\bar{z'}\rangle  &=& e^{-\frac{|z|^{2} + |z'|^{2}}{2}}\sum_{n,l=0}^{\infty}
\frac{z^{n}\bar{z'}^{l}}{\sqrt{n !l !}}|\Psi_{nl}\rangle \otimes |\alpha_{k}\rangle\cr
&=& U_{1}(z)U_{2}(z')|\Psi_{00}\rangle \otimes 
|\alpha_{k}\rangle.
\end{eqnarray}
\epro

{\bf Proof.} The proof is straightforward by combining the actions of both unitary operators $ U_{1}(z)$ and $U_{2}(z')$ on the state $|\Psi_{00}\rangle $, respectively.
$\hfill{\square}$

\brmk
Besides, dealing wth the following unitary transformation,
\beq\label{unitransf000}
\mathcal V : L^{2}(\C, \frac{d\bar{z} \wedge d{z}}{\imath}) &\longrightarrow& L^{2}(\C, d\nu(\bar{ z}, z))\cr
\Psi(\bar{z}, z) &\longmapsto& (U\Psi)(\bar{z}, z) = \sqrt{\pi} e^{| z|^{2}}\Psi(\bar{z}, z),
\eeq
where $d\nu(\bar{z}, z) = \frac{e^{-|z|^{2}}}{2\pi}\frac{d\bar{ z} \wedge d{ z}}{\imath}$, the operators defined in (\ref{fock005}) become
\beq{\label{operat00}}
\mathcal  A_{+} &=:&  \mathcal VA_{+}\mathcal V^{-1}  = \partial_{\bar{ z}} 
= A_{+} - z, \; \mathcal  A^{*}_{+} =:\mathcal VA^{*}_{+}\mathcal V^{-1}  =  -\partial_{ z }  +  2\bar{ z}
= A^{*}_{+} + \bar{ z},\cr
\mathcal  A_{-} &=:&  \mathcal VA_{-}\mathcal V^{-1}  =  \partial_{ z }  
= A_{-} - \bar{z}, \; \mathcal  A^{*}_{-} = \mathcal VA^{*}_{-}\mathcal V^{-1} =  -\partial_{\bar{ z}} + 2z  
= A^{*}_{-} +  z,
\eeq
\beq{\label{operat03}}
\tilde{\mathcal A}_{+} &=:&  \mathcal V\tilde A_{+}\mathcal V^{-1} =   \partial_{\bar{ z}} 
= \tilde A_{+} -  z, \; \tilde{\mathcal A}^{*}_{+} =:  \mathcal V\tilde A^{*}_{+}\mathcal V^{-1} = -\partial_{ z }  
= \tilde A^{*}_{+} + \bar{ z},\cr
\tilde{\mathcal A}_{-} &=:&  \mathcal V\tilde A_{-}\mathcal V^{-1} =\partial_{ z } 
= \tilde A_{-} - \bar{ z}, \; \tilde{\mathcal A}^{*}_{-} =:  \mathcal V\tilde A^{*}_{-}\mathcal V^{-1} = -\partial_{\bar{ z}} + 2 z 
= \tilde A^{*}_{-} +  z.
\eeq
Moreover, take the antiunitary operator defined in \cite{abh} as 
\beq
J: \mathcal B_{2}(\mathfrak H) \longrightarrow \mathcal B_{2}(\mathfrak H), \quad J(|\phi\rangle\langle \psi|) = |\psi\rangle\langle \phi|, \quad  \forall\phi, \psi \in \mathfrak H.
\eeq
The CSs given by $U_{1}(z)U_{2}(z')|\Psi_{00}\rangle$ in (\ref{bcs000}), using (\ref{operat00}),  correspond to the bi-coherent states, denoted $\eta^{\rm{bcs}}_{\bar u, v}$, given by
\beq
\eta^{\rm{bcs}}_{\bar u, v} = e^{\frac{|u|^2 +|v|^2}{2}}e^{\bar u \mathcal  A^{*}_{-} - u \mathcal  A_{-}}e^{v \mathcal  A^{*}_{+} - \bar v \mathcal  A_{+}}H_{00}, \quad u, v \in \C
\eeq
associated to the Hamiltonian $\mathcal H^\uparrow - \mathcal H^\downarrow \equiv H_{2_{OSC}} - \hat H_{1_{OSC}}$, with $\mathcal H^\uparrow = \mathcal V H_+\mathcal V^{-1}$ and $\mathcal H^\downarrow  = \mathcal V H_-\mathcal V^{-1}$, $\{H_+, H_-\}$ provided in (\ref{form01})-(\ref{form02}).  $H_{00}$ refers to the Hermite polynomials. These latters, verified with $\mathcal J = (\mathcal{V}\circ \mathcal{W})J(\mathcal{V}\circ \mathcal{W})^{-1}$, $\mathcal{W}$ and  $\mathcal{V}$ given in (\ref{map1}) and (\ref{unitransf000}), the following relations (see \cite{abh}):
\beq
\mathcal J\eta^{\rm{bcs}}_{\bar u, v} = \eta^{\rm{bcs}}_{\bar v, u}, \qquad \mathcal JH^\uparrow\mathcal J = \mathcal H^\downarrow
\eeq
implying physically that the map $\mathcal J$ reverses the uniform magnetic field, from $\overrightarrow B$ to $-\overrightarrow B$, thus transforming the Hamiltonian $\mathcal H^\uparrow = \mathcal V H_+\mathcal V^{-1}$ to $\mathcal H^\downarrow  = \mathcal V H_-\mathcal V^{-1}$.

\ermk

\brmk
The operators $U_{1}$ and $U_{2}$ realize unitary irreducible representations of the Weyl-Heisenberg 
group $G_{W-H}$ on the separable Hilbert space $\mathfrak H_{s} = L^{2}(\R^{2}, dxdy)$. This group has for  generic operators \cite{ali-antoine-gazeau} 
$U(\vartheta, q, p) = e^{i (\vartheta I  + p Q - q P)}$. Indeed, with $\vartheta = 0$, we get: 
\begin{eqnarray}
U_{1}(z) = e^{z b'^{\dag} - \bar{z}b'} = e^{i (pQ'_{1} - qP'_{1})}, \;
U_{2}(z') = e^{z' \mathfrak b'^{\dag} - \bar{z'}\mathfrak b'} = e^{i (p'Q'_{2} - q'P'_{2})} ,
\end{eqnarray}
with $z = \frac{q- i p}{\sqrt{2}}, \, z'= \frac{q'- i p'}{\sqrt{2}}$, 
where the position and momentum operators are $Q'_{1} = \frac{1}{\sqrt{2}}(b' + b'^{\dag}), P'_{1}= \frac{i}{\sqrt{2}}(b'^{\dag}-b')$ and $Q'_{2} = \frac{1}{\sqrt{2}}(\mathfrak b' + \mathfrak b'^{\dag}), P'_{2} = \frac{i}{\sqrt{2}}(\mathfrak b'^{\dag}-\mathfrak b')$, respectively, 
acting on $\mathfrak H_{s} = L^{2}(\R^{2}, dxdy)$.
\ermk

\section{Concluding remarks}\label{sec5}

In this work, the Wigner transform has been first explored on the quantum Hamiltonians,  with both discrete and continuous spectra,  describing the motion of an electron in an electromagnetic field. Using this map defined from $\mathcal B_2(\mathfrak{H})$ to $L^2(\R, dxdy)$, where $\mathfrak{H} = L^2(\R)$,  as key ingredient, the counterparts of the discrete
and continuous parts of the physical system Hamiltonians have been  provided on 
$L^2(\R) \otimes L^2(\R)$ and $L^2(\R)$, respectively. Then, coherent states 
have been constructed and  satisfy the Gazeau-Klauder
coherent states criteria that are the continuity in the labels, the resolution of the identity, the action identity,  and the temporal
stability. Moreover, using unitary operators
realizing unitary irreducible representations of the Weyl-Heisenberg group, coherent states of same type have been also achieved. The discussion can be extended when dealing with the statistical behavior of the constructed coherent states by defining the denstity operators associated to both disrete and continuous spectra Hamiltonians by following \cite{aremuajmp}. These aspects will be in the core of a forthcoming  work.

\vspace{5mm}

\section*{Appendix} 
Proof of Proposition \ref{propcs001}.

From the definition, we have
 \begin{eqnarray}
 &&\| |J,\gamma; J', \gamma';l; K_{1}, \theta_{1}\rangle - |\tilde J,\tilde \gamma; \tilde J', \tilde \gamma';l; \tilde K_{1}, \tilde \theta_{1}\rangle  \|^2 \cr
 &=& \| |J, \gamma; J', \gamma';l\rangle \otimes |K_{1}, \theta_{1}\rangle - |\tilde J, \tilde \gamma; \tilde J', \tilde \gamma';l\rangle \otimes|\tilde K_{1}, \tilde \theta_{1}\rangle  \|^2 \cr
 &=& \left\{\langle J, \gamma; J', \gamma';l| \otimes \langle K_{1}, \theta_{1}| - \langle \tilde J, \tilde \gamma; \tilde J', \tilde \gamma';l| \otimes \langle \tilde K_{1}, \tilde \theta_{1}|\right\}\cr
 &&\times\left\{|J, \gamma; J', \gamma';l\rangle \otimes |K_{1}, \theta_{1}\rangle - |\tilde J, \tilde \gamma; \tilde J', \tilde \gamma';l\rangle \otimes|\tilde K_{1}, \tilde \theta_{1}\rangle\right\}\cr
 &=& 1- \left\{\sum^{\infty}_{l=0}\sum^{\infty}_{m,n=0}\frac{J'^{l/2}e^{i l \gamma'}}{\left[\mathcal N(J) \mathcal N(J')\right]}\frac{J^{m/2}}{\sqrt{m !l !}}
\frac{\tilde J'^{l/2}e^{i l \tilde \gamma'}}{\left[\mathcal N(\tilde J) \mathcal N(\tilde J')\right]}\frac{\tilde J^{n/2}}{\sqrt{n !l !}}e^{i m \gamma}e^{-i n \tilde \gamma}\delta_{mn}\right\}\cr
 && \otimes \left\{[(N(K_{1}))(N(\tilde K_{1}))]^{-1} \int_{0}^{\infty}\int_{0}^{\infty}
\frac{K_{1}^{E_{\alpha_{1}}}\tilde K_{1}^{ E_{\alpha'_{1}}}}{e^{\frac{1}{2}\beta (E^{2}_{\alpha_{1}} + E^{2}_{\alpha'_{1}})}}e^{i \theta_{1} (E_{\alpha_{1}})}e^{-i \tilde \theta_{1} ( E_{\alpha'_{1}})} \delta(\alpha'_{1}-\alpha_{1}) dE_{\alpha'_{1}}dE_{\alpha_{1}}\right\}\cr
 \cr
 &&-\left\{\sum^{\infty}_{l=0}\sum^{\infty}_{m,n=0}\frac{\tilde J'^{l/2}e^{i l \tilde \gamma}}{\left[\mathcal N(\tilde J) \mathcal N(\tilde J')\right]}\frac{\tilde J^{m/2}}{\sqrt{m !l !}}
\frac{J'^{l/2}}{\left[\mathcal N(J) \mathcal N(J')\right]}\frac{J^{n/2}}{\sqrt{n !l !}} e^{i m \tilde \gamma}e^{-i n \gamma}\delta_{mn}\right\}\cr
 && \otimes \left\{[(N(K_{1}))(N(\tilde K_{1}))]^{-1} \int_{0}^{\infty}\int_{0}^{\infty}
\frac{K_{1}^{E_{\alpha_{1}}}\tilde K_{1}^{ E_{\alpha'_{1}}}}{e^{\frac{1}{2}\beta (E^{2}_{\alpha_{1}} + E^{2}_{\alpha'_{1}})}}e^{-i \theta_{1} (E_{\alpha_{1}})}e^{+i \tilde\theta_{1} ( E_{\alpha'_{1}})}\delta(\alpha_{1}-\alpha'_{1}) dE_{\alpha'_{1}}dE_{\alpha_{1}}\right\}\cr
\cr
&& +1. \nonumber \\  
\end{eqnarray}
Thereby 
\begin{equation}
\lim_{(J,\gamma; J', \gamma';l; K_{1}, \theta_{1})
\rightarrow (\tilde J,\tilde \gamma; \tilde J', \tilde \gamma';l; \tilde K_{1}, \tilde \theta_{1})}\| |J,\gamma; J', \gamma';l; K_{1}, \theta_{1}\rangle - |\tilde J,\tilde \gamma; \tilde J', \tilde \gamma';l; \tilde K_{1}, \tilde \theta_{1}\rangle \|^2
=0,
\end{equation}
which completes the proof.

$\hfill{\square}$

{ Proof of Proposition \ref{propcs007}. 
 
 From (\ref{vcs00}), one has
\begin{eqnarray}
&&|J, \gamma;J', \gamma';l;K_{1}, \theta_{1}\rangle \langle J, \gamma;J', \gamma';l;K_{1}, \theta_{1}| \cr
&&= \frac{J'^{l}}{\left[\mathcal N(J) \mathcal N(J')\right]^2}\sum^{\infty}_{n,p=0}\frac{J^{(n + p)/2}e^{-i (n-p) \gamma}}{\sqrt{n !l !p !l !}}|n;l\rangle \langle p,l| \otimes N(K_{1})^{-2}\cr
&&\otimes \int_{0}^{\infty}\int_{0}^{\infty}
\frac{K_{1}^{E_{\alpha_{1}} + E_{\alpha'_{1}}}}{e^{\frac{1}{2}\beta (E^{2}_{\alpha_{1}} + E^{2}_{\alpha'_{1}})}}e^{-i \theta_{1} (E_{\alpha_{1}} - E_{\alpha'_{1}})}| \alpha'_{1}\rangle \langle\alpha_{1}| dE_{\alpha'_{1}}dE_{\alpha_{1}}.
\end{eqnarray}
Then,
\begin{eqnarray}
&&\int_{-\infty}^{\infty}\int_{0}^{2\pi}\int_{0}^{2\pi}|J, \gamma;J', \gamma';l;K_{1}, \theta_{1}\rangle \langle J, \gamma;J', \gamma';l;K_{1}, \theta_{1}|\mathcal N(J)^{2}
 \mathcal N(J')^{2} N(K_{1})^{2}\cr
&& \frac{d\gamma}{2\pi}\frac{d\gamma'}{2\pi}\frac{d\theta_{1}}{2\pi} \cr
&&=J'^{l}\sum^{\infty}_{n,p=0}\frac{J^{(n + p)/2}}{\sqrt{n !l !p !l !}}\delta_{np}|n,l\rangle \langle p,l| \otimes \
\int_{0}^{\infty}\int_{0}^{\infty}
\frac{K_{1}^{E_{\alpha_{1}} + E_{\alpha'_{1}}}}{e^{\frac{1}{2}\beta (E^{2}_{\alpha_{1}} + E^{2}_{\alpha'_{1}})}}\delta(\alpha'_{1}-\alpha_{1})|\alpha'_{1}\rangle \langle\alpha_{1}| dE_{\alpha'_{1}}dE_{\alpha_{1}}  \cr
\cr
&&=J'^{l}\sum^{\infty}_{n=0}\frac{J^{n}}{n !}|n,l\rangle \langle n,l| \otimes \
\int_{0}^{\infty}
\frac{K_{1}^{2E_{\alpha_{1}}}}{e^{\beta E^{2}_{\alpha_{1}}}}|\alpha_{1}\rangle \langle\alpha_{1}| dE_{\alpha_{1}}.\nonumber\\
\end{eqnarray}
such that
\begin{eqnarray}
&&\int^{\infty}_{-\infty}
\int^{\infty}_{0}\int^{\infty}_{0}
\int^{\infty}_{0}\int_{0}^{2\pi}\int_{0}^{2\pi}
|J,\gamma; J', \gamma';l; K_{1}, \theta_{1}\rangle \langle J,\gamma; J', \gamma';l; K_{1}, \theta_{1}| \cr
&& \frac{d\gamma}{2\pi}\frac{d\gamma'}{2\pi}\frac{d\theta_{1}}{2\pi}\mathcal N(J)^{2}
\mathcal N(J')^{2}
 N(K_{1})^{2}d\nu(J)d\nu(J')d\rho(K_{1})\cr
&& = \sum^{\infty}_{n=0}\int_{0}^{\infty}\int_{0}^{\infty}\frac{J^{n}}{n !}d\nu(J)\frac{J'^{l}}{l !}d\nu(J')|n,l\rangle \langle n,l| \otimes \
\int_{0}^{\infty}
\frac{K_{1}^{2E_{\alpha_{1}}}}{e^{\beta E^{2}_{\alpha_{1}}}}d\rho(K_{1})| \alpha_{1}\rangle \langle\alpha_{1}| dE_{\alpha_{1}}.\nonumber 
\end{eqnarray}
The measures $d\nu(J), d\nu(J')$  and $\sigma(K_{1})$ with expressions
\begin{eqnarray}
d\nu(J) = e^{-J} dJ, \quad d\nu(J') = e^{-J'} dJ',  \quad \sigma(K_{1}) = \frac{1}{K_{1}\sqrt{\beta \pi}}e^{-\frac{1}{\beta}}(\ln K_{1})^{2}\backslash
\end{eqnarray}
are such that the moment problems given by
\begin{eqnarray}
\int^{\infty}_{0}J^{n}d\nu(J) = n !, \; \int^{\infty}_{0}J'^{l}d\nu(J') = l !, \; \int_{0}^{\infty}
K_{1}^{2E_{\alpha_{1}}} \sigma(K_{1})dK_{1} = e^{\beta E^{2}_{\alpha_{1}}}
\end{eqnarray}
are satisfied. Therefore,
\begin{eqnarray}
&&\int^{\infty}_{-\infty}
\int^{\infty}_{0}\int^{\infty}_{0}
\int^{\infty}_{0}\int_{0}^{2\pi}\int_{0}^{2\pi}
|J,\gamma; J', \gamma';l; K_{1}, \theta_{1}\rangle \langle J,\gamma; J', \gamma';l; K_{1}, \theta_{1}| \cr
&& \frac{d\gamma}{2\pi}\frac{d\gamma'}{2\pi}\frac{d\theta_{1}}{2\pi}\mathcal N(J)^{2}
\mathcal N(J')^{2}
 N(K_{1})^{2}d\nu(J)d\nu(J')d\rho(K_{1})\cr
&& = \sum^{\infty}_{n=0}|n,l\rangle \langle n,l| \otimes \
\int_{0}^{\infty}|\alpha_{1}\rangle \langle\alpha_{1}| dE_{\alpha_{1}}   = I_{\mathfrak H^l_D}\otimes I_{\mathfrak H_C}.
\end{eqnarray}
$\hfill{\square}$
}

Proof of  Proposition \ref{propcs003}.

By definition, we have  
\begin{eqnarray}
e^{-i \mathcal H_{1} t}|J,\gamma; J', \gamma';l;K_{1}, \theta_{1}\rangle 
&=& e^{-i \mathcal H_{1_{D}}t}|J,\gamma; J', \gamma';l\rangle \otimes e^{-i {\mathcal H}_{1_{C}}t}|K_{1}, \theta_{1}\rangle\cr
&=& \left[\mathcal N(J) \mathcal N(J')\right]^{-1}J'^{l/2}e^{i l \gamma'}\sum^{\infty}_{n=0}\frac{J^{n/2}e^{-i n (\gamma + \omega_{c}t)}}{\sqrt{n ! l !}}|n,l\rangle \cr 
&&\otimes \left [N(K_{1})\right]^{-1}  \int_{0}^{\infty}\frac{K_{1}^{E_{\alpha_{1}}}}{e^{\frac{1}{2}\beta E^{2}_{\alpha_{1}}}}e^{-i (\theta_{1} + t) E_{\alpha_{1}}}|\alpha_{1}\rangle dE_{\alpha_{1}}\cr\cr
&=& |J,\gamma + \omega_{c}t;J', \gamma';l;K_{1}, \theta_{1} + t\rangle.\nonumber\\
\end{eqnarray}

$\hfill{\square}$

Proof of Proposition  \ref{propcs005}. 

We have 
\begin{eqnarray}
&&\langle J,\gamma; J', \gamma';l; K_{1}, \theta_{1}|{\mathcal H}_{1_{D}} + { {\mathcal H}}_{1_{C}}|J,\gamma; J', \gamma';l; K_{1}, \theta_{1}\rangle \cr
&=& \langle  J,\gamma; J', \gamma';l| \otimes \langle K_{1}, \theta_{1}|{\mathcal H}_{1_{D}} + { {\mathcal H}}_{1_{C}}|J,\gamma; J', \gamma';l\rangle \otimes | K_{1}, \theta_{1}\rangle\cr
&=& \langle J,\gamma; J', \gamma';l| {\mathcal H}_{1_{D}} |J,\gamma; J', \gamma';l\rangle  + \langle K_{1}, \theta_{1}|{ {\mathcal H}}_{1_{C}}|K_{1}, \theta_{1}\rangle
\end{eqnarray}
where 
{
\begin{eqnarray}
&&\sum^{\infty}_{l=0}\langle  J,\gamma; J', \gamma';l; K_{1}, \theta_{1}| {\mathcal H}_{1_{D}} | J,\gamma; J', \gamma';l; K_{1}, \theta_{1}\rangle \cr
&=&\sum^{\infty}_{l=0}\langle J,\gamma; J', \gamma';l| {\mathcal H}_{1_{D}} |J,\gamma; J', \gamma';l\rangle \cr
&=& \sum^{\infty}_{l=0}\frac{J'^{l}}{\left[\mathcal N(J) \mathcal N(J')\right]^2}\sum^{\infty}_{n, m=0}\frac{J^{n/2 + m/2}e^{-i (n -m)\gamma}\omega_{c}n}{\sqrt{n !m !}l !}\delta_{nm}\cr
&=& \omega_{c} J\left[\sum^{\infty}_{l=0}\frac{J'^{l}}{\left[\mathcal N(J) \mathcal N(J')\right]^2}\sum^{\infty}_{n=1}\frac{J^{n-1}}{(n -1)!l !}\right]\cr
&=& \omega_{c} J \nonumber\\
\end{eqnarray}
}
and  
\begin{eqnarray}
 &&\langle J,\gamma; J', \gamma';l; K_{1}, \theta_{1}|{ {\mathcal H}}_{1_{C}}|J,\gamma; J', \gamma';l; K_{1}, \theta_{1}\rangle  \cr
&=&\langle K_{1}, \theta_{1}|{ {\mathcal H}}_{1_{C}}|K_{1}, \theta_{1}\rangle \cr
&=& N(K_{1})^{-2}\int_{0}^{\infty}\int_{0}^{\infty}
\frac{K_{1}^{E_{\alpha_{1}} + E_{\alpha'_{1}}}}{e^{\frac{1}{2}\beta (E^{2}_{\alpha_{1}} + E^{2}_{\alpha'_{1}})}}e^{-i \theta_{1} (E_{\alpha_{1}} - E_{\alpha'_{1}})}\delta(\alpha'_{1} - \alpha_{1})E_{\alpha_{1}} dE_{\alpha'_{1}}dE_{\alpha_{1}}  \cr
\cr
&=& N(K_{1})^{-2}\int_{0}^{\infty}\frac{K_{1}^{2E_{\alpha_{1}}}}{e^{\beta E^{2}_{\alpha_{1}}}}E_{\alpha_{1}}dE_{\alpha_{1}}
\end{eqnarray}
implying
\begin{eqnarray}
 \langle K_{1}, \theta_{1}|{ {\mathcal H}}_{1_{C}}|K_{1}, \theta_{1}\rangle  &=& N(K_{1})^{-2}\int_{0}^{\infty}\frac{K_{1}^{2E_{\alpha_{1}}}}{e^{\beta E^{2}_{\alpha_{1}}}}E_{\alpha_{1}}dE_{\alpha_{1}}: = {{\mathcal H}}_{1_{C}}(K_{1})
\end{eqnarray}
where the new action variable ${ {\mathcal H}}_{1_{C}}(K_{1})=:\mathcal J(K_{1})$ is assumed to be invertible versus $K_1$ as  in \cite{ben-klauder}. Assuming that under some condition reached by a strictly
increasing or decreasing function ${ {\mathcal H}}_{1_{C}}(K_{1})$ (${ {\mathcal H}}_{1_{C}}(K_{1}) > 0$ or ${ {\mathcal H}}_{1_{C}}(K_{1}) < 0$) such that $K_1(\mathcal J)$ can be
determined, then the CSs $|\mathcal J, \theta_1\rangle := |K_1(\mathcal J), \theta_1\rangle$ fulfill the action identity property: $\langle \mathcal J, \theta_1|{ {\mathcal H}}_{1_{C}}|\mathcal J, \theta_1 \rangle = \langle K_1(\mathcal J), \theta_1|{{\mathcal H}}_{1_{C}}|K_1(\mathcal J), \theta_1 \rangle = \mathcal J$.

$\hfill{\square}$

Proof of the  Proposition \ref{unitopcss000}.

From the completeness relation in the $\{|x\rangle\}$ representation $\displaystyle\int |x\rangle \langle x| dx = I_{\mathfrak H_C}  $,  one has
\begin{eqnarray}
|z,\bar{z'};l;K,\theta;\beta\rangle  &=& f(K,\theta)|z,\bar{z'};l\rangle  + 
e^{-i \beta}g(z, \bar{z'})|K,\theta\rangle \cr
          &=& f(K,\theta)e^{-\frac{|z'|^{2} + 
|z|^{2}}{2}}\frac{\bar{z'}^{l}}{\sqrt{l 
!}}\sum^{\infty}_{n=0}\frac{z^{n}}{\sqrt{n !}}\frac{(b^{\dag})^{n}}{\sqrt{(2 m\omega_c \hbar)^{n} n !}}|\Psi_{0l}\rangle  \cr
&& + e^{-i \beta}g(z, \bar{z'})
\mathcal N_{\rho}(K)^{-\frac{1}{2}}\int^{\infty}_{0}
\frac{K^{\epsilon^{-}_{\alpha}/2}e^{i \epsilon_{\alpha}\theta}}{\sqrt{\rho(\epsilon^{-}_{\alpha})}}d\epsilon^{-}_{\alpha} \int |x\rangle dx \langle x|\alpha\rangle \cr
&=& f(K,\theta)e^{-\frac{|z'|^{2} + 
|z|^{2}}{2}}\frac{\bar{z'}^{l}}{\sqrt{l 
!}}\sum^{\infty}_{n=0}\frac{z^{n}}{\sqrt{n !}}\frac{(b'^{\dag})^{n}}{\sqrt{n !}}|\Psi_{0l}\rangle 
\cr
&& + e^{-i \beta}g(z, \bar{z'})
\mathcal N_{\rho}(K)^{-\frac{1}{2}}\int^{\infty}_{0}
\frac{K^{\epsilon^{-}_{\alpha}/2}e^{i \epsilon_{\alpha}\theta}}{\sqrt{\rho(\epsilon^{-}_{\alpha})}}d\epsilon^{-}_{\alpha} \int |x\rangle dx \phi_{\alpha}(x)  \cr
&=& f(K,\theta)e^{-\frac{|z'|^{2} + 
|z|^{2}}{2}}\frac{\bar{z'}^{l}}{\sqrt{l 
!}}\sum^{\infty}_{n=0}\frac{z^{n}}{n !}(b'^{\dag})^{n}|\Psi_{0l}\rangle \cr
&& + e^{-i \beta}g(z, \bar{z'})
\mathcal N_{\rho}(K)^{-\frac{1}{2}}\int^{\infty}_{0}
\frac{K^{\epsilon^{-}_{\alpha}/2}e^{i \epsilon_{\alpha}\theta}}{\sqrt{\rho(\epsilon^{-}_{\alpha})}}d\epsilon^{-}_{\alpha} \int |x\rangle dx D_{\varepsilon}\phi_{\alpha}(x + \varepsilon)  \cr
&=& f(K,\theta)e^{-\frac{|z'|^{2}}{2}}\frac{\bar{z'}^{l}}
{\sqrt{l !}}U_{1}(z)|\Psi_{0l}\rangle 
\cr
&&  + e^{-i \beta}g(z, \bar{z'})
\mathcal N_{\rho}(K)^{-\frac{1}{2}}\int^{\infty}_{0}
\frac{K^{\epsilon^{-}_{\alpha}/2}e^{i \epsilon_{\alpha}\theta}}{\sqrt{\rho(\epsilon^{-}_{\alpha})}}d\epsilon^{-}_{\alpha} \int |x\rangle dx D_{\varepsilon}\phi_{\alpha}(x + \varepsilon) \nonumber
\\
\end{eqnarray}
which completes the proof.

$\hfill{\square}$

\vspace{5mm}


\end{document}